\documentstyle[12pt,aaspp4]{article}
\eqsecnum
\slugcomment{Submitted to {\it The Astrophysical Journal}}

\begin{document}

\lefthead{Najita, Tiede, and Carr }
\righthead{Low-Mass IMF in IC348}

\title{From Stars to Super-planets: the Low-Mass IMF in the 
Young Cluster IC348
\footnote{Based on observations with the NASA/ESA Hubble Space Telescope, 
obtained at the Space Telescope Science Institute, which is operated 
by the Association of Universities for Research in Astronomy, Inc., 
under NASA contract No. NAS5-26555.}}

\author{Joan R. Najita, Glenn P.  Tiede} 
\affil{National Optical Astronomy Observatories, 950 N. Cherry Ave., Tucson, AZ 85719}
\authoremail{\{tiede, najita\}@noao.edu}
\author{and}
\author{John S. Carr}
\affil{Naval Research Laboratory, 4555 Overlook Ave., Washington, D.C. 
20375-5320}
\authoremail{carr@mriga.nrl.navy.mil}

\begin{abstract}

We investigate the low-mass population of the young cluster IC348
down to the deuterium--burning limit, a fiducial boundary between 
brown dwarf and planetary mass objects, using a new and innovative method for the 
spectral classification of late-type objects.  
Using photometric indices, constructed from HST/NICMOS
narrow-band imaging, that measure the strength of the $1.9\micron$
water band, we determine the spectral type and reddening
for every M-type star in the field, thereby separating cluster members
from the interloper population.  
Due to the efficiency of our spectral classification technique,
our study is complete from $\sim 0.7 M_\odot$ to $0.015~M_\odot$.
The mass function derived for the cluster in this interval,
$dN/d\log M \propto M^{0.5}$, is similar to that obtained for the Pleiades,
but appears significantly more abundant in brown dwarfs than the mass
function for companions to nearby sun-like stars.  
This provides compelling observational evidence for different formation and
evolutionary histories for substellar objects formed in isolation 
vs. as companions.  Because our determination of the IMF is complete 
to very low masses, we can place interesting constraints on the role of 
physical processes such as fragmentation in the star and planet formation
process and the fraction of dark matter in the Galactic halo that 
resides in substellar objects.

\end{abstract}


\section{Introduction}

The low-mass end of the stellar initial mass function (IMF) 
is of interest for our understanding of both 
baryonic dark matter in the Galaxy and, perhaps more importantly,  
the formation processes governing stars, brown dwarfs, and planets. 
In the stellar mass regime, the complex interplay between a wide 
array of physical processes is believed to determine the eventual 
outcome of the star formation process, the masses of stars. 
These diverse processes include those that govern molecular cloud 
structure and evolution, subsequent gravitational collapse, 
disk accretion, stellar winds, multiplicity, and stellar mergers.
What is the distribution of object masses that results from 
the interaction between these processes?
Do the same processes that form stars also produce less massive 
objects extending into the brown dwarf and planetary regimes?
While such questions can be answered directly by constructing 
inventories of stellar and substellar objects, 
there is also the hope that the same set of data can 
shed light on the nature of the interaction between the physical processes 
and, thereby, bring us closer to a predictive theory of star 
and brown dwarf formation.  

While the stellar IMF has long been studied (e.g., \cite{sal55}), 
the very low-mass and substellar IMF is much less well known 
since the very existence 
of substellar objects has only recently been demonstrated, and reliable 
inventories of substellar objects are only now becoming available. 
The Pleiades has proven to be one of the most popular sites for 
low-mass IMF studies both due to its proximity ($d\sim 125$ pc) 
and because it is at an age ($\sim 100$ Myr) at which our understanding 
of stellar evolution is fairly robust.  The large area subtended by 
the Pleiades poses several challenges: studies of the low mass IMF 
must survey large areas and distinguish low mass cluster 
members from the growing Galactic interloper population at faint 
magnitudes.  For example, recent deep imaging surveys of the Pleiades
carried out over several square degrees have used broad band color
selection criteria to probe the cluster IMF to masses below the
hydrogen burning limit (e.g., to $\sim 0.04 M_\odot$; Bouvier et al.\ 1999), 
where the fraction of objects that are cluster members is much less 
than 1\%.

In a complementary development, new large area surveys 
(e.g., 2MASS, DENIS, and SDSS) are now probing the low mass IMF of the field 
population in the solar neighborhood, extending into the substellar regime. 
In an account of the progress to date, Reid et al.\ (1999) model 
the spectral type distribution of the low mass population drawn 
from 2MASS and DENIS samples obtained over several hundred square degrees 
in order to constrain the low mass IMF.  
Since substellar objects cool as they age, the observed spectral type 
distribution depends on both the mass and age distributions of the local 
field population.  As a result, the 
lack of strong constraints on the age distribution poses a challenge 
for the determination of the field IMF at low masses. 
For example, assuming a flat age distribution over $0-10$ Gyr, 
Reid et al. find an IMF that is fairly flat, 
$dN/d\log M \propto M^{\alpha}$ where $\alpha \sim$ -1 to 0, 
where the uncertainty in the slope does not include the uncertainty 
in the age distribution of the population.

In comparison with the solar neighborhood and 
older open clusters such as the Pleiades, 
young stellar clusters ($\lesssim 10$ Myr) are a 
complementary and advantageous environment in 
which to carry out low-mass IMF studies.  
As in the situation for the Pleiades, stars in young clusters
share a common distance and metallicity and, at low masses,
are much brighter due to their youth.  As a well recognized
consequence, it is possible to readily detect and study even
objects much below the hydrogen-burning limit.
In addition, young clusters also offer some significant advantages 
over the older open clusters.
For example, since young clusters are less dynamically evolved 
than older open clusters, the effects of mass segregation and the
evaporation of low mass cluster members are less severe.
Since young clusters are less dynamically evolved, they 
also subtend a more compact region on the sky.  As a result, 
the fractional foreground and background contamination is much
reduced and reasonable stellar population statistics can be
obtained by surveying small regions of the sky.
These advantages are (of course) accompanied by challenges 
associated with the study of young environments.  These include 
the need to correct for both differential reddening toward individual 
stars and infrared excess, the excess continuum emission that is 
believed to arise from circumstellar disks.  
Pre-main sequence evolutionary tracks pose the greatest challenge to
the interpretation of the observations because the tracks have little
observational verification, especially at low masses and young ages. 
The temperature calibration for low-mass pre-main-sequence stars is 
an additional uncertainty.

While thus far the luminosity advantage of young clusters has 
been used with  
great success to detect some very low mass cluster members  
(e.g., $\lesssim 0.02 M_\odot$ objects in IC348 [Luhman 1999] 
and the $\sigma$~Ori cluster [Zapatero Osorio et al.\ 2000]), 
attempts to study the low mass IMF in young clusters have stalled 
at much higher masses,
in the vicinity of the hydrogen burning limit (e.g., the Orion 
Nebula Cluster---Hillenbrand 1997), due to the need for complete 
sampling to low masses and potentially large extinctions.  
Since reddening and IR excesses can greatly complicate the determination 
of stellar masses from broad band photometry alone (e.g., Meyer et al.\ 1997), 
stellar spectral classification to faint magnitudes, an often 
time-consuming task, is typically required. 

Stellar spectral classification in young clusters has
been carried out using a variety of spectroscopic methods.  
These include the use of narrow atomic and molecular features
in the $K$-band (e.g., \cite{ali95}; \cite{gre95}; Luhman et al.\ 1998, 
hereinafter LRLL), the $H$-band (e.g., \cite{mey96}),
and the $I$-band (e.g., \cite{hil97}), each of which have their advantages.
While spectral classification at the longer wavelengths is better 
able to penetrate higher extinctions,
spectral classification at the shorter wavelengths
is less affected by infrared excess. 
With the use of high spectral resolution and the availability of
multiple stellar spectral features, it is possible to diagnose and
correct for infrared excess.  This technique has been used with great
success at optical wavelengths in the study of T Tauri star
photospheres (e.g., \cite{har89}).
Alternatively, the difficulty of correcting for infrared excess
can be avoided to a large extent by studying
somewhat older (5-20 Myr old) clusters, 
in which infrared excesses are largely absent but significant dynamical
evolution has not yet occurred.

In this paper, we develop an alternative, efficient method of spectral
classification: filter photometric measures of water absorption 
band strength as an indicator of stellar spectral type.
Water bands dominate the infrared spectra of M stars and are highly 
temperature sensitive, increasing in strength with decreasing 
effective temperature down to the coolest M dwarfs known 
($\sim 2000$K; e.g., \cite{jon94}).  
The strength of the water bands and their rapid variation 
with effective temperature, in principle, allows the precise 
measurement of spectral type from moderate signal-to-noise photometry. 
At the same time, water bands are relatively insensitive to gravity 
(e.g., \cite{jon95}),
particularly above 3000K, becoming more sensitive at lower 
temperatures where dust formation is an added complication  
(e.g., the Ames-Dusty models; \cite{all00}; \cite{all98b}). 
Synthetic atmospheres (e.g., NextGen: \cite{hau99}; \cite{all97}) also indicate 
a modest dependence of water band strength on metallicity 
(e.g., \cite{jon95}).

Because strong absorption by water in the Earth's atmosphere
can complicate the ground-based measurement of the depth of 
water bands, we used HST NICMOS filter photometry to
carry out the measurements.
The breadth of the water absorption bands requires that any measure of
band strength adequately account for the effects of reddening.
Consequently, we used a 3 filter system to construct a reddening 
independent index that measures the band strength.
Of the filters available with NICMOS, only the narrow band
F166N, F190N, and F215N filters which sample the depth of 
the 1.9~$\mu$m water band proved suitable. 
On the one hand, the narrow filter widths had the advantages of
excluding possible stellar or nebular line emission and
limiting the differential reddening across the bandpass.
On the other hand,
similar filters with broader band passes would have made it
feasible to study much fainter sources, e.g., in richer clusters 
at much larger distances.
Despite the latter difficulty, there were suitable nearby clusters
such as IC348 to which this technique could be profitably applied.

IC348 is a compact, young cluster located near an edge of the Perseus
molecular cloud.  It has a significant history of optical study (see,
e.g., \cite{her98} for a review), and because of its proximity ($d\sim
300$pc), youth ($< 10$ Myr), and rich, compact nature, both the star formation
history and the mass function that characterizes the cluster have been
the subject of several recent studies.

Ground-based $J$, $H$, $K$ imaging of the cluster complete to $K$=14 
(\cite{lad95}) revealed signficant spatial structure, 
in which the richest stellar grouping 
is the ``a'' subcluster ($r=3.5'$; hereinafter IC348a) with 
approximately half of the cluster members.  
The near-IR colors indicate that  IC348 is an advantageous environment 
in which to study the stellar properties of a young cluster since 
only a moderate fraction of cluster members possess near-IR 
excesses ($\sim$20\% for the cluster overall;  
$\sim$ 12\% for IC348a) 
and most cluster members suffer moderate extinction
($A_V \sim 5$ with a spread to $A_V$ $>$ 20).  
Lada \& Lada (1995) showed that the 
$K$-band luminosity function of IC348 is consistent with 
a history of continuous star formation over the last $5-7$ Myr 
and a time-independent Miller-Scalo IMF in the mass range $0.1-20 M_\odot$. 
The inferred mean age of a few Myr is generally consistent with the lack of a 
significant population of excess sources since disks are believed
to disperse on a comparable timescale (\cite{mey00}).

Herbig (1998) subsequently confirmed a significant age spread to 
the cluster ($0.7-12$ Myr) based on BVRI imaging of a 
$\sim 7\arcmin \times 12\arcmin$ region, which included much of 
IC348a, and $R$-band spectroscopy of a subset of sources in the field.
In the mass range in which the study is complete ($M_* > 0.3 M_\odot$), 
the mass function slope was found to be consistent with that of 
\cite{sca86}.
A more detailed study of a $5\arcmin \times 5\arcmin $ region centered 
on IC348a was carried out by \cite{luh98} using 
IR and optical spectroscopy complete to $K=12.5$. 
They also found an age spread to the subcluster ($5-10$ Myr), 
a mean age of $\sim 3$ Myr, 
and evidence for a substellar population.
The mass function of the subcluster was found to 
be consistent with \cite{mil79} in the mass range $0.25-3 M_\odot$ 
(i.e., flatter in slope than deduced by Herbig)  
and flatter than Miller-Scalo at masses below $0.25 M_\odot$; 
however, completeness corrections were significant 
below $\sim 0.1 M_\odot$.
\cite{luh99} has further probed the substellar population of IC348
using optical spectral classification of additional sources 
($I\lesssim 19.5$) both in and beyond the $5^\prime \times 5^\prime$ core. 

In this paper, we extend previous studies of IC348 by probing 
4 magnitudes below the $K$ spectral completeness limit of \cite{luh98},
enabling a more detailed look at the population in the low-mass stellar and 
substellar regimes.  We find that, with our spectral classification 
technique, our measurement of the IMF in IC348 is complete to the 
deuterium burning limit ($\sim 0.015 M_\odot$), a fiducial boundary 
between brown dwarf and planetary mass objects (e.g., Saumon et al.\ 2000; 
Zapatero Osorio et al.\ 2000).  To avoid potential misunderstanding, 
we note that this boundary is only very approximate. 
A precise division between the brown dwarf 
and planetary regimes is unavailable and perhaps unattainable 
in the near future given the current disagreement over fundamental issues 
regarding the definition of the term ``planet''.  These include 
whether the distinction between brown dwarfs and planets should be 
made in terms of mass or formation history (e.g., gravitational 
collapse vs.\ accumulation) and 
whether planetary mass objects that are not companions can even be 
considered to be ``planets''. 
Here, we hope to side-step such a discussion at the outset 
and, instead, explore how the IMF of isolated objects over the range 
from $\sim  1 M_\odot$ to $\sim 0.015 M_\odot$, 
once measured, can advance the discussion, i.e., provide clues to 
the formation and evolutionary histories of stellar and substellar objects. 
The HST observations are presented in section 2. 
The resulting astrometry and near-infrared luminosity functions 
are discussed in sections 3 and 4.  
In section 5, we discuss the calibration of the water index and 
the determination of stellar spectral types.
The reddening corrections are discussed in section 6, and the 
resulting observational HR diagram in section 7. 
In section 8, we identify the interloper population and compare 
the cluster population with the predictions of pre-main sequence 
evolutionary tracks. 
Given these results, in section 9, we 
identify possible cluster binaries and 
derive a mass function for the cluster. 
Finally, in section 10, we present our conclusions.

\section{Observations, Data Reduction, and Calibration}

\subsection{Photometry}

\begin{figure}
\figurenum{1}
\label{finder}
\plotfiddle{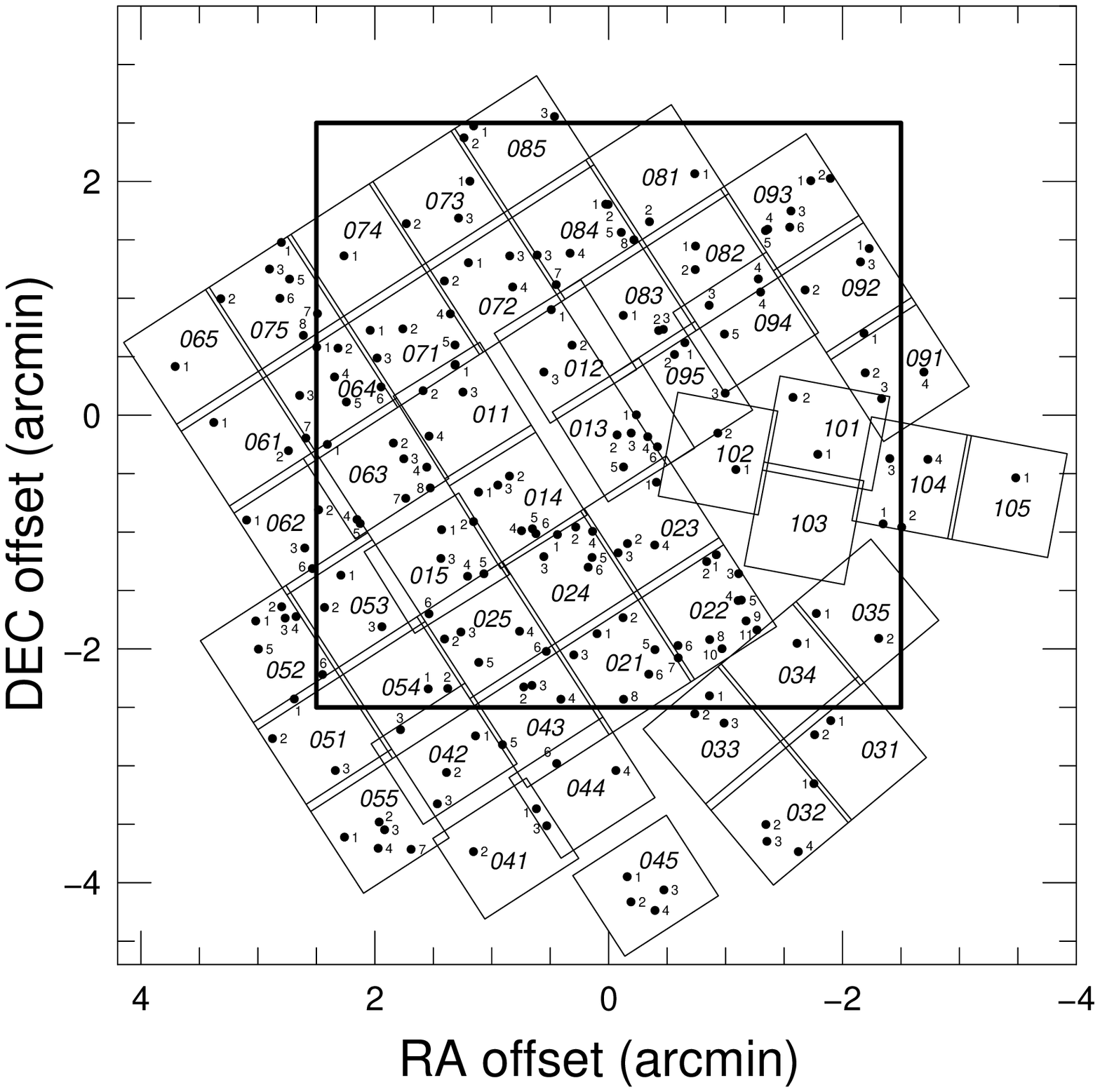}
{6.0in}{0}{75}{75}{-300}{10}
\caption{Finder chart showing the relative positions of the HST/NIC3
fields and the detected objects.  
The axes indicate offsets in arcminutes from the nominal
center of IC348a, $\alpha=3^{\rm h}44^{\rm m}31\fs9$, 
$\delta=32^\circ 09\arcmin 54\farcs2$ (J2000).  The large square
(heavy line) is the $5\arcmin \times 5\arcmin$ cluster core defined by
\cite{luh98}.  Our 3-digit field designations (large numbers) and the
stellar designations in each field (small numbers) are also shown.  The
bright stars ($K\lesssim 9$) that were intentionally excluded in
positioning the fields are not shown.
}
\end{figure}

We obtained HST NIC3 narrow band photometry for 50 $(51'' \times 51'')$ 
fields in the IC348a subcluster, nominally centered at 
$\alpha=3^{\rm h}44^{\rm m}31\fs9$, 
$\delta=32^\circ 09\arcmin 54\farcs2$ (J2000).  
The NICMOS instrument and its on-orbit performance have been
described by \cite{tho98} and \cite{cal98}.  
Figure~\ref{finder}
shows the relative positions of the fields with respect
to the $5\arcmin \times 5\arcmin$ core of the subcluster. 
The NIC3 field positions were 
chosen to avoid bright stars much above the saturation limit ($K\lesssim 9$) 
and to maximize area coverage.  As a result, the fields 
are largely non-overlapping, covering most of the 
$5\arcmin \times 5\arcmin$ core and 
a total area of $34.76$ sq.\ arcmin.
Each field was imaged in the narrow band F166N, F190N, and F215N
filters, centered at 1.66 $\mu$m, 1.90 $\mu$m, 2.15 $\mu$m
respectively, at two dither positions separated by $5.1\arcsec$.
The exposure time at each dither position was 128 seconds, obtained through
four reads of the NIC3 array in the SPARS64 MULTIACCUM sequence, for
a total exposure time in each field of 256 seconds.

To calibrate the non-standard NIC3 colors, we observed a set of
23 standard stars chosen to cover spectral types K2 through M9
that have the kinematics and/or colors typical of solar neighborhood disk 
stars (e.g., \cite{leg92}; see Table~\ref{standards}) and, therefore, 
are likely to have metallicities similar to that of the cluster stars.
Although most of the standard stars were main-sequence dwarfs, we also
observed a few pre-main sequence stars in order to explore the effect
of lower gravity.
We chose for this purpose pre-main sequence stars known to have low
infrared excesses (weak lined T Tauri stars; WTTS) so that the observed
flux would be dominated by the stellar photosphere.
The standard stars were observed in each of the F166N, F190N, and F215N 
filters and with the G141 and G206 grisms.
The stars were observed with each spectral element at two or three 
dither positions separated by $5.1\arcsec$ in MULTIACCUM mode.  

Since NICMOS does not have a shutter, 
the bright standard stars could potentially saturate the array
as the NIC3 filter wheel rotates through the broad or intermediate band
filters located between the narrow band filters and grisms
used in the program.  To avoid the
resulting persistence image that would compromise the photometric accuracy,
dummy exposures, taken at a position offset from where the science
exposure would be made, were inserted between the science exposures in
order to position the filter wheel at the desired spectral element before
actually taking the science exposure.

Much of the data for IC348 (45 of the 50 fields) and all of the
data for the standard stars were obtained during the first 
(January 12 -- February 1, 1998) and second (June 4 -- 28, 1998) NIC3 
campaigns in which the HST secondary was moved to bring NIC3 into focus.
A log of our observations is provided in Table~\ref{log}.  
The data were processed through the usual NICMOS {\it calnic} pipeline 
(version 3.2) with the addition of one step. 
After the cosmic ray identification, column bias offsets were removed 
from the final readout in order to eliminate the ``banding'' 
(constant, incremental offsets of 
$\sim 30$ counts about 40 columns wide)
present in the raw data. 

No residual reflection nebulosity is noticeable in the reduced 
(dither-subtracted) images. 
Consequently, removal of nebular emission was not a 
concern for the stellar photometry.
To perform the stellar photometry, we first identified sources in each
of the images using the IRAF routine {\it daofind}.  Due to the strongly
varying noise characteristics of the NIC3 array, {\it daofind} erroneously 
identified numerous noise peaks as point sources, and so the detections 
were inspected frame by frame to eliminate spurious detections.  
A detection was considered to be real if the source was detected in both the 
F215N and F190N frames.  With these identification criteria, we were 
likely to obtain robust detections of heavily extincted objects 
(in F215N) as well as spectral types for all identified sources, 
F190N typically having the lowest flux level at late spectral types. 

Since the frames are sparsely populated, we used the aperture photometry 
routine {\it phot} to measure the flux of each identified source.  
To optimize the signal-to-noise of the photometry on faint 
objects ($K\gtrsim 16$), we adopted a 4-pixel radius photometric 
aperture that included the core of the PSF and $\sim 91\%$ of the 
total point source flux (the exact value varied by about $1\%$ from 
filter to filter) with an uncertainty in the aperture correction of 
$<1\%$ in all filters.  The aperture correction was derived from 
observations of calibration standards and/or bright, unsaturated objects 
in the IC348 fields. 
Despite the difference in focus conditions between the data taken 
in and out of the NIC3 campaigns, the aperture corrections were  
statistically identical.  As a result, the same aperture and 
procedures were used for both data sets.  
The conversion from ADU/s to both Janskys and magnitudes 
was made using the photometric constants kindly provided 
by M. Rieke (1999, personal communication).  These constants
are tabulated in Table~\ref{photcal}. 

\subsection{Spectroscopy}

In order to confirm the calibration of the filter photometric 
water index against stellar spectral type, we also obtained 
NIC3 G141 and G206 grism spectra for 17 of our 23 standard stars. 
The spectral images were processed identically to the photometric 
images, including the removal of the bias jumps.  
The spectra were extracted using NICMOSlook (version 2.6.5; \cite{pir98a}), 
the interactive version of the standard pipeline tool 
(CalnicC; \cite{pir98b}) 
for the extraction of NIC3 grism spectra. 
The details of the extraction process and subsequent analysis are presented 
in Tiede et al.\ (2000).
The 1.9 $\mu$m H$_2$O band strengths obtained from a preliminary 
analysis of the spectra were found to be consistent with the 
filter photometric results reported in section 5.

\subsection{Intrapixel Sensitivity and Photometric Accuracy}

Because infrared arrays may have sensitivity variations at the sub-pixel
scale, the detected flux from an object,
when measured with an undersampled PSF,
may depend sensitively on the precise position of the object within in a
pixel.
As shown by \cite{lau99}, such intrapixel sensitivity effects can be significant
when working with undersampled NIC3 data ($0.2\arcsec$ pixels).
To help us quantify the impact of this effect on our data set,
Lauer kindly calculated for us the expected intrapixel
dependence of the detected flux from a point source as a function of
intrapixel position,
using TinyTim PSFs appropriate for the filters in our study
and the NIC3 intrapixel response function deduced in \cite{lau99}.
As expected, the intrapixel sensitivity effect is more severe at
shorter wavelengths where the undersampling is more extreme.
In the F215N filter, the effect is negligible: the variation in the
detected flux as a function of intrapixel position is within
$\pm 0.3$\% of the flux that would be detected with a well sampled PSF.
For the F190N and F166N filters, the same quantity varies within $\pm 3.5$\%
and $\pm 8.5$\%, respectively.

Although intrapixel sensitivity can be severe at the shorter
wavelengths, the effect on photometric colors is mitigated
if the intrapixel response is similar for the three filters
(the assumption made here) and the sub-pixel positional offsets
between the observations in each filter are small.
For example, with no positional offset between the 3 filters, the
error in the
reddening independent water index, $Q_{\rm H2O}$, discussed in section 5,
is $<$1\% which impacts negligibly on our conclusions.
Since pointing with HST is expected to be accurate to better than
a few milliarcseconds for the $\sim 17$ minute duration of the
observations on a given cluster field (M. Lallo 1999, personal communication),
pointing drifts are unlikely to introduce significant positional
offsets.  The HST jitter data for our observations confirm the
expected pointing accuracy.  Over the $\sim 5$ minute duration of the
exposure in a single filter, the RMS pointing error is on average
$\sim 4$ milliarcseconds (0.02 NIC3 pixels).

Systematic positional offsets between filters could also arise from
differing geometric transformations between the filters. To test this,
we examined the centroid position of the bright cluster sources and
standard stars for individual dither positions in each filter.
No systematic differences in centroid positions between filters
were found.
The 1--$\sigma$ scatter about the mean was 0.05 pixels which represents
the combination of our centroiding accuracy and any true positional
variations.  To quantify the impact of the latter possibility on our results,
random positional variations of 0.05 pixels in each filter translate
into a maximal error in $Q_{\rm H2O}$ of less than $\pm$4\%.

\section{Astrometry}

\begin{figure}
\figurenum{2}
\label{areas}
\plotfiddle{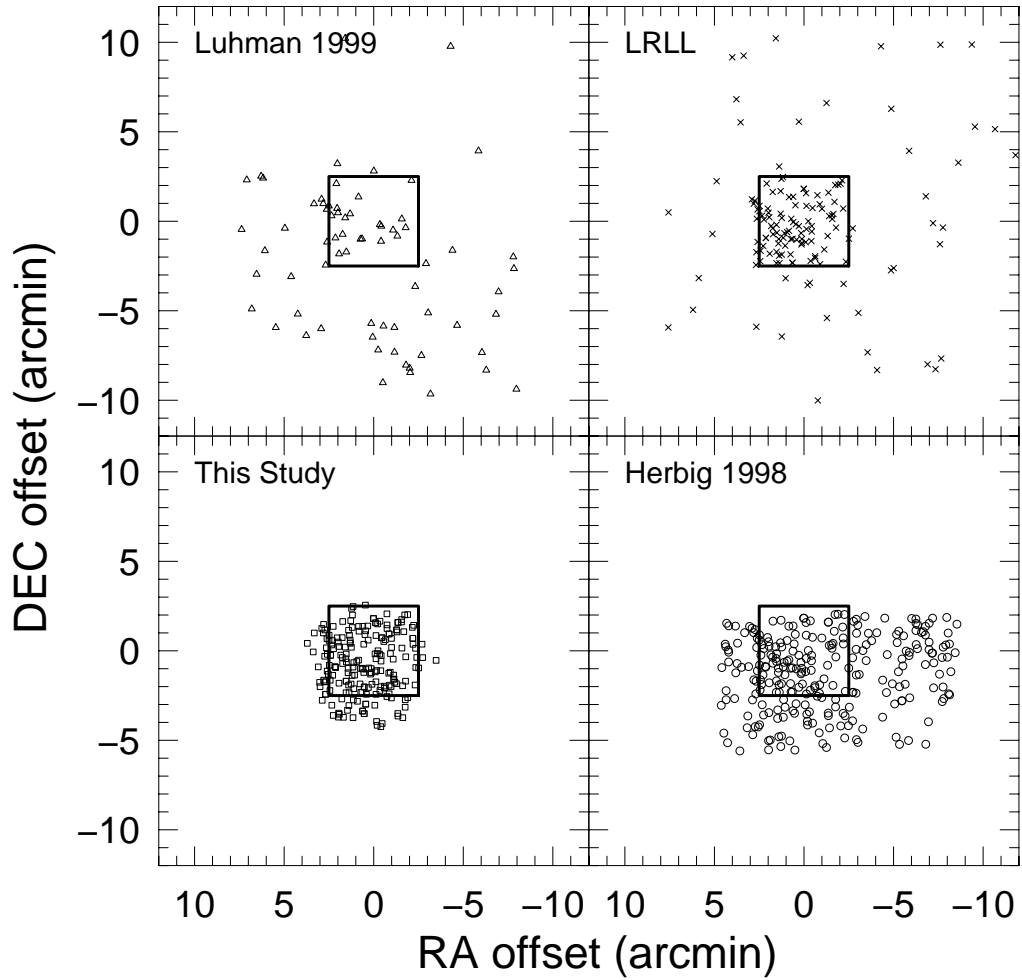}
{6.0in}{0}{75}{75}{-300}{30}
\caption{Spatial distribution of the stellar samples from recent
studies of IC348 including the present study.  The relative positions
of each sample with respect to the $5\arcmin \times 5\arcmin$
core of IC348a (heavy-lined square) are shown.
The axes indicate offsets in arcminutes from the nominal center of
the subcluster.}
\end{figure}

Because three of the recent studies of IC348 
(\cite{her98}, \cite{luh98}, and Luhman 1999) 
have examined regions surrounding and including IC348a,  
we can directly compare the previous results with ours via the 
overlaps in the stellar samples.  
Figure~\ref{areas} shows the spatial distribution of the samples 
from the previous and present studies.
The present study covers a more compact region than the previous studies, 
but is complete to much greater depth. 

Table~\ref{astrometry} presents the source designations for all of the stars
in our sample, the corresponding designations from previous studies, 
and the J2000 celestial coordinates of each 
star.  Our designations are comprised of the 3-digit field number
followed by the 2-digit number of the star in that field.  For example, 
021-05 is from field 021 and is star number 5 in that field.  
The celestial coordinates in Table~\ref{astrometry} are based on the
NICMOS header values associated with the central pixel in each field.
The total error in the relative accuracy of the coordinates 
due to photometric centroiding, geometric field distortion, and repeat pointing errors, 
are estimated to be $\lesssim 0 \farcs 2$ per star.  
This error is a function of the stellar position in the NIC3 field of view:  
stars located toward the corners of a frame have larger errors primarily due to 
field distortion which we have not attempted to correct.  
While absolute astrometry is not required for the present study, 
we can obtain an estimate of the absolute astrometric error by comparing 
our coordinates to those obtained in previous investigations. 
Comparison with the celestial coordinates reported in \cite{luh98} typically 
resulted in disagreements of less than $1\arcsec$.

\section{Completeness and Luminosity Functions}

\subsection{Completeness and Photometric Accuracy}

At the bright end, our sample is limited by saturation. 
Inspection of the error flags output by CALNICA implied that 
our saturation limits are 
$10.96 \pm 0.49$ magnitudes in F166N, 
$10.89 \pm 0.44$ in F190N; and 
$10.62 \pm 0.35$ in F215N.  
The flux range over which saturation occured reflects 
the sensitivity variation across the array and 
the variation in the intrapixel position of individual stars.

Given the noise characteristics of, 
and significant quantum efficiency variations across, the NIC3 array, 
we used simulated data to evaluate 
the efficiency of our detection algorithm at the faint end and the 
accuracy of our photometric measurements.
We first added to a representative frame for each filter
a known number of point sources, positioned randomly within the frame, 
with known magnitudes and zero color,
then performed detection and stellar photometry on the frames in a method
identical to those used for the real data.  Since crowding was
not an issue in the real frames, care was taken to ensure that none 
of the artificial stars where lost to superposition.
While we did not explore the full color 
range of the actual data set, the adopted simulation was sufficient to 
obtain a robust estimate of our detection efficiency in the individual 
filters. 
 
Artificial PSFs were generated using the program 
TinyTim version 4.4 (\cite{kri97}).  
Each artificial PSF was 
created with a factor of 10 oversampling, i.e, in a 240 $\times$ 240 grid 
with each element of the grid representing $0.02\arcsec$ on the sky, to 
facilitate sub-pixel interpolation in positioning the artificial stars.  
The extent of the artificial PSF ($2.4\arcsec$) was chosen to equal the 
radius at which the flux level for even the brightest stars in the 
data set is less than the noise fluctuations in the background.

Inspection of the empirical luminosity functions, the theoretical
photometric errors, and signal-to-noise values indicated that our
sample was likely complete to $\sim 17.5$ mag (0.1 mJy in F215N).  To
derive the completeness limit quantitatively for each band, we
created two sets of artificial stars to be added and recovered from a
representative frame in each band.  The first set of 50 stars was linearly
distributed over the magnitude range in which photometric errors
become significant (15.0 to 19.5).  The second set of 50 stars was
linearly distributed between 17.0 and 18.5 magnitude in order to
``zero-in'' on the $100\%$ completeness limit.  After the addition of
the artificial stars with the appropriate noise, each of the images 
was photometrically processed in a manner identical to the real data frames.

\begin{figure}
\figurenum{3}
\label{com1}
\plotfiddle{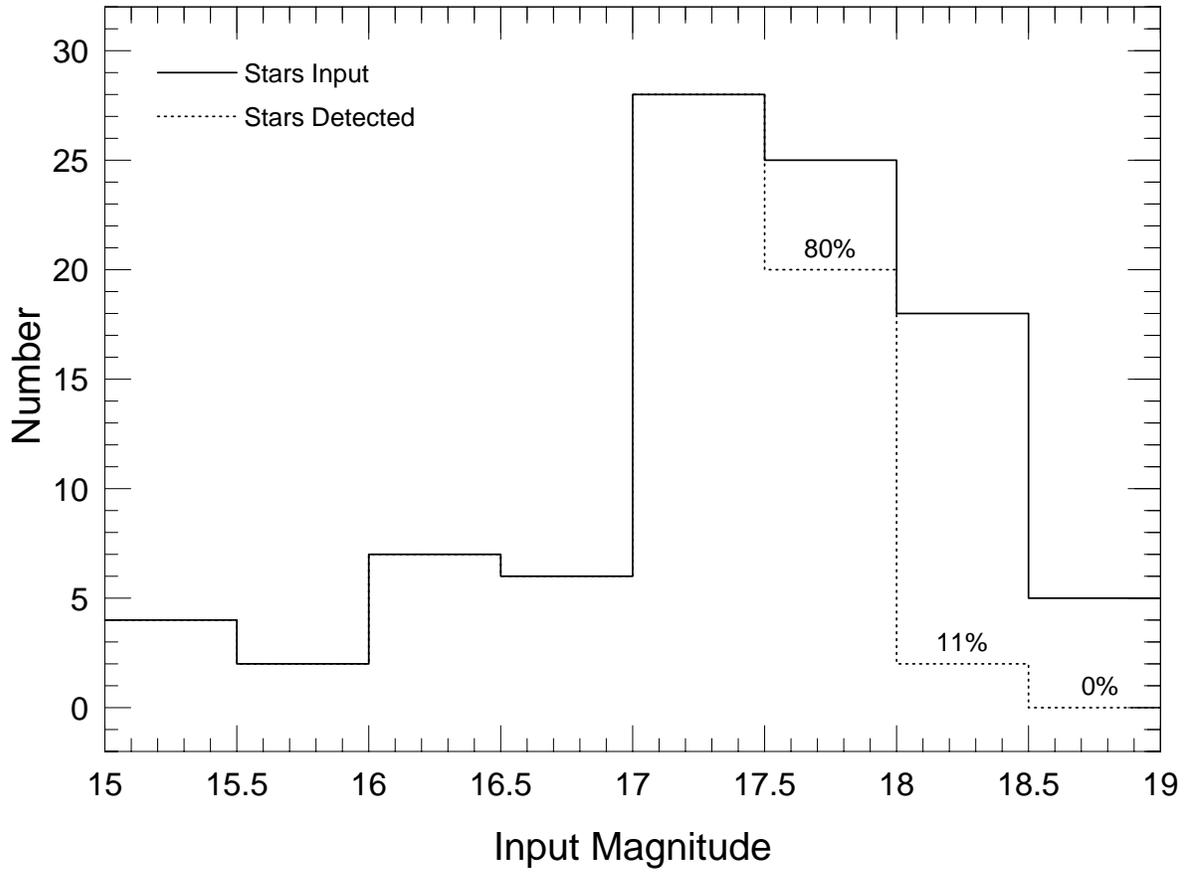}
{5.0in}{0}{65}{65}{-250}{0}
\caption{Comparison of the number of artificial stars input (solid line) 
and detected (dotted line) in F215N. The artificial stars are recovered 
with 100\% efficiency to 17.5 magnitude.  The completeness fractions at 
fainter magnitudes are as noted.}
\end{figure}

The completeness as a function of F215N magnitude is displayed in 
Fig.~\ref{com1}. 
The results are essentially identical for F190N.
The Figure shows the number of stars input into (solid line) and 
the number detected in (dotted line) each 0.5 magnitude bin.  
Our photometry is $100\%$ complete 
through the bin centered at 17.25 magnitudes, beyond which   
the detection efficiency drops rapidly.  It is 80\% at 17.75, 
11\% at 18.25, and finally no detections beyond 18.5.  
When the results are tabulated in 0.1 magnitude bins, 
we find that we are $100\%$ complete to 17.6 magnitudes.
Since the last $100\%$ complete bin only contains 5 stars and because
the rest of analysis is done in 0.5 magnitude increments, we adopt 
17.5 magnitudes as a conservative estimate of our $100\%$ completeness limit.

\begin{figure}
\figurenum{4}
\label{err1}
\plotfiddle{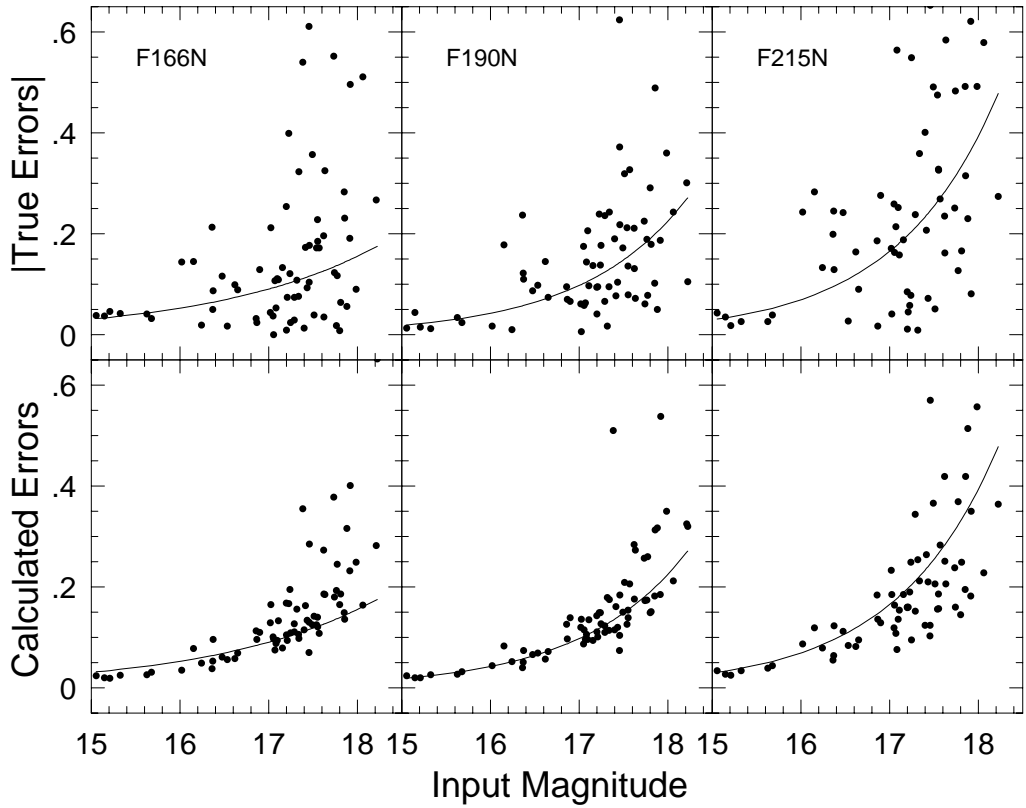}
{5.0in}{0}{60}{60}{-250}{0}
\caption{Photometric accuracy as a function of magnitude.  The top panels
show the absolute value of the ``True Errors'' (measured magnitude - input
magnitude) measured from simulated data.  The curves in the top panels are 
exponential fits to the values for the individual bands.  The bottom panels 
show the estimated errors (see text) for the same data.  The curves
are the same as in the top panels.  
}
\end{figure}

In addition to calculating the completeness limit, the artificial stars
also allowed us to gauge the accuracy of our photometry and photometric 
error estimates.  Since we knew the magnitudes of the
artificial stars that we added to the frame, we could calculate the ``True
Error'' of each photometric measurement (True Error $\equiv$ measured 
magnitude $-$ input magnitude).  The top panels of Fig.~\ref{err1} show 
the absolute value of the resulting true errors as a function of  
input magnitude.  
For each photometric measurement, we calculated the photometric uncertainty 
due to photon statistics.  
The bottom panels of Fig.~\ref{err1} show this estimated error versus 
input magnitude.  Although the
scatter in the absolute value of the true errors is much larger than the
scatter in the estimated errors, the estimated errors provide a good 
approximation to the true errors in an average sense.
This remains true down to the completeness limit. 
In all three bands, the estimated errors fall along the curves fit to the 
true errors with significant deviation only below $\sim 17.5$ magnitudes. 

\subsection{Empirical and Combined Luminosity Functions}

\begin{figure}
\figurenum{5}
\label{lfs}
\plotfiddle{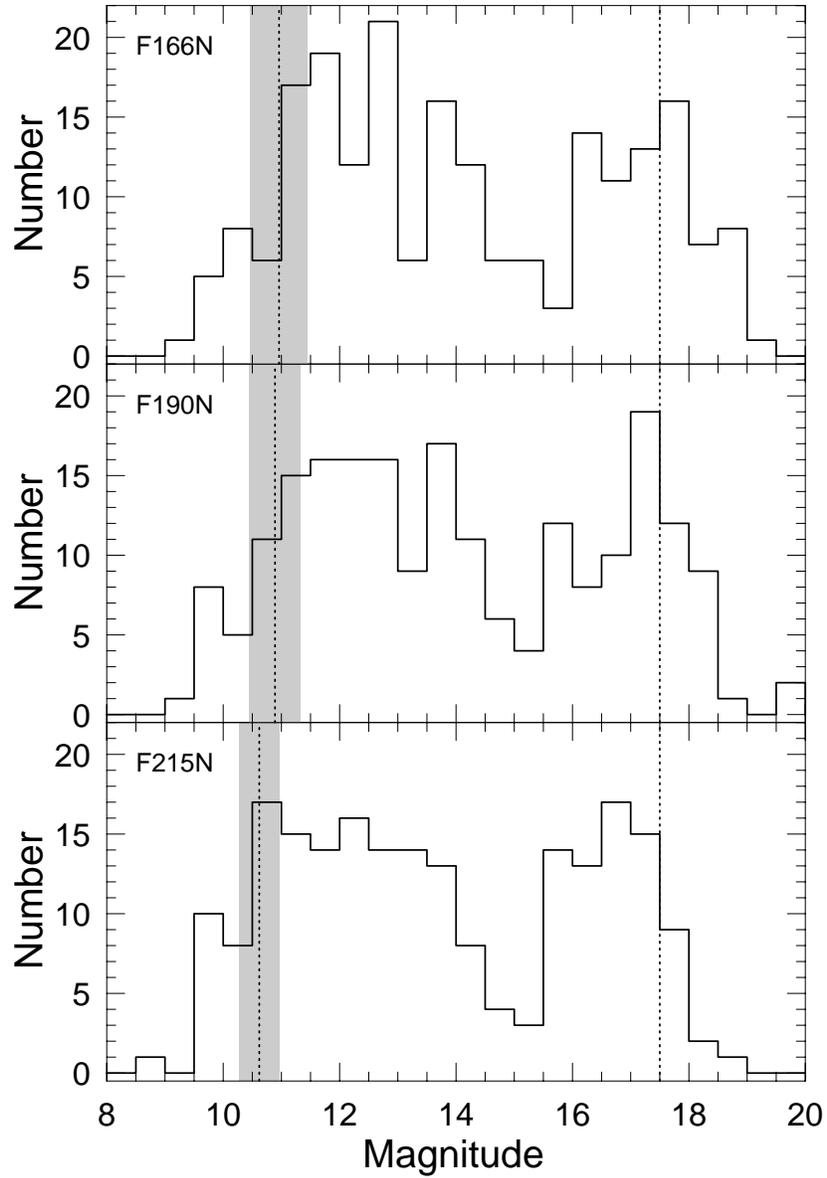}
{6.5in}{0}{65}{65}{-210}{0}
\caption{Luminosity functions for each of the narrow band filters.  
No correction for reddening or completeness has been made.
The vertical dotted lines indicate the mean saturation and 
photometric completeness limits in each filter.  Saturation occurs over 
a range in magnitude (grey bar) due to pixelization and variations in 
flat field reponse across the detector.
}
\end{figure}

The luminosity functions (LFs) for each of the narrow band filters are shown
in Fig.~\ref{lfs}.  No corrections for reddening or completeness have
been made.  The range in magnitude over which saturation occurs
is indicated by the grey band in each panel.  The vertical dotted lines 
indicate the mean saturation limit and the completeness limit of 
$17.5$ magnitudes.  The F215N luminosity function is 
relatively flat between the saturation and completeness limits, 
with a dip between 14 and 15.5 magnitudes.  The structure in the 
F166N and F190N luminosity functions is similar.

In order to compare our LF with previously determined LFs for IC348, we 
converted our measured F215N magnitudes to standard $K$ magnitudes.
The F215N filter measures a relatively feature-free 
region of the standard CTIO/CIT $K$ filter.  Therefore, 
the F215N magnitude should correlate well with $K$, requiring a 
zero-point offset and possibly, due to increasing water band
strengths in the coolest M stars, a color term.
To determine the offset, we compared our F215N magnitudes with 
published $K$ magnitudes for the 61 stars in our sample that are in 
common with Lada \& Lada (1995; see tabulation in \cite{luh98}) 
and/or \cite{luh99} and are below the 
saturation limit ($K>11$).  The fit
had a slope statistically identical to unity ($1.003 \pm 0.012$), so
we derived the mean offset between the two magnitude systems,
$\langle(K-F215N)\rangle = -0.115 \pm 0.011$, where the error is the
error in the mean.  The 1--$\sigma$ residual to the fit was $0.085$.
This residual is comparable to the typical combined photometric accuracy of
the Luhman and our data.  
We investigated a possible color term in
the transformation, but found that if any is present it is smaller than
this scatter about the mean. 

The accuracy and completeness of the bright end of our 
luminosity function ($K\lesssim 11$) is compromised by both saturation 
and our deliberate avoidance of bright cluster stars.  To correct for this
deficiency, we combined our derived $K$ photometry at $K\ge 11$ 
with $K$ photometry of the $5\arcmin\times 5\arcmin$ core from \cite{luh98} 
for $K< 11$.  
This combination is reasonable since, as shown in Fig.~\ref{finder}, 
the region of our survey largely overlaps the 
$5\arcmin\times 5\arcmin$ core. 
To correct for the different areas covered by two surveys, 
we multiplied the counts in each bin of 
the LRLL luminosity function by the ratio of the survey areas, 
$34.76/25.00 = 1.39$.  

\begin{figure}
\figurenum{6}
\label{klf}
\plotfiddle{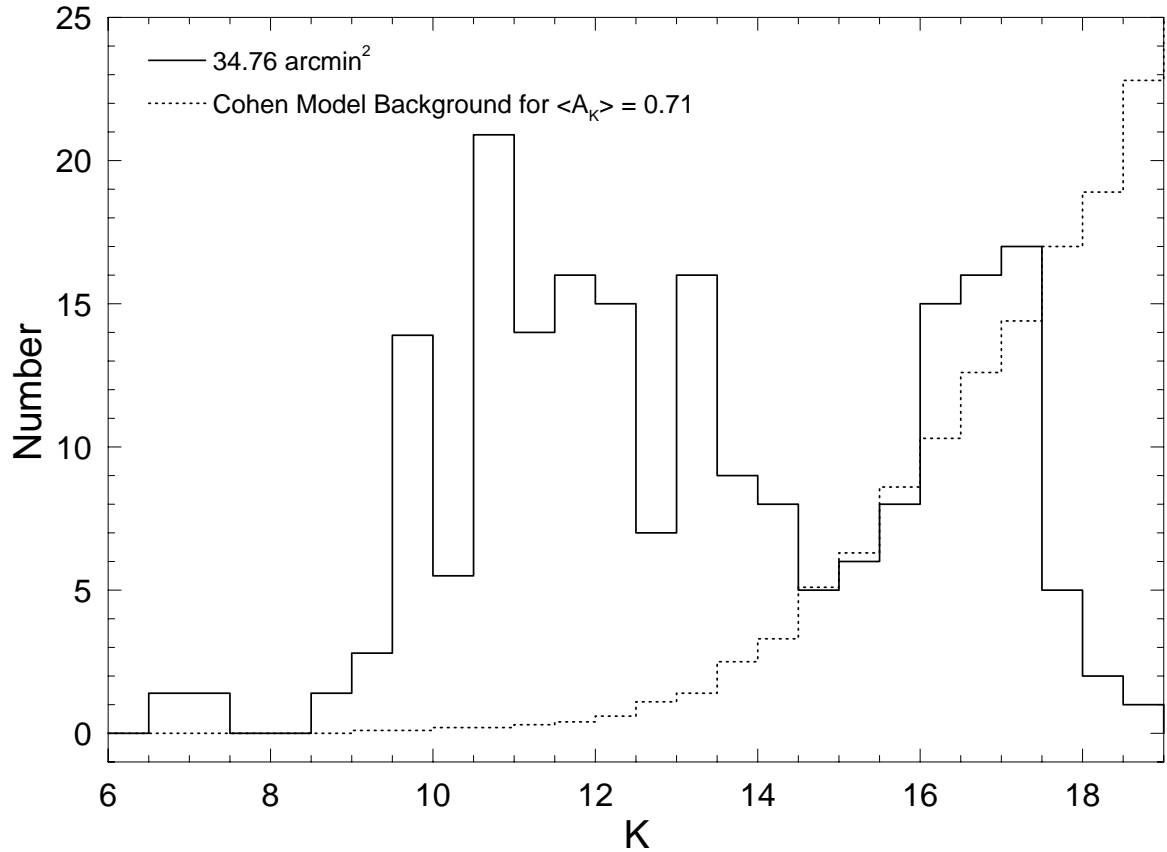}
{6.0in}{0}{65}{65}{-245}{0}
\caption{The combined $K$ luminosity function for the IC348a fields 
(solid line histogram).  The luminosity function, 
complete to $K \simeq 17.5$,  combines photometry from 
\cite{luh98} at magnitudes above our saturation limit ($K = 11.0$) with 
our derived $K$ photometry at fainter magnitudes.  
The dotted histogram shows an estimate, based on  
the star count model of \cite{coh94}, 
of the background contamination to the IC348a fields. 
}
\end{figure}

The combined $K$ luminosity function for our 34.76 sq. arcmin region, 
complete to $K \simeq 17.5$, is shown in Fig.~\ref{klf} 
as the solid line histogram.  
To estimate the background contribution to the $K$ luminosity function, 
we used the prediction of the star count model of \cite{coh94}.  
The predicted background $K$ counts, 
reddened by the mean reddening of the background population 
($A_K=0.71$; see section 6), is shown as the dotted line 
histogram in Fig.~\ref{klf}.
In section 8, we compare in greater detail the results for our data set
with the predictions of the model.
Here we simply note a few points.  The contamination of the 
cluster by background stars is insignificant to $K \approx 13$ and the 
number of cluster stars is larger than the number of background stars until 
the $K = 14.25$ bin.  While the background rises steadily, we appear to 
have detected a few cluster stars to our completeness limit.

\section{Spectral Classification}

To derive spectral types for the stars in the sample, 
we combined the measured narrow band fluxes into a reddening 
independent index, 
\[
Q_{\rm H2O} \equiv -2.5\log{\left(\frac{\rm F166}{\rm F190}\right)} + 1.37 \times 
2.5\log{\left(\frac{\rm F190}{\rm F215}\right)}, 
\]
that measures the strength of the $1.9~\mu$m H$_2$O absorption band. 
In this expression, F166, F190, and F215 are the fluxes in the 
F166N, F190N, and F215N filters, respectively. 
The value 1.37 is the ratio of the reddening color excesses:
\[
\frac{E({\rm F166/F190})}{E({\rm F190/F215})} = 1.37,
\]
which is derived from the infrared extinction law 
$A_\lambda/A_V = 0.412(\lambda/\mu {\rm m})^{-1.75}$
(\cite{tok99}). 

\begin{figure}
\figurenum{7}
\label{qfit}
\plotfiddle{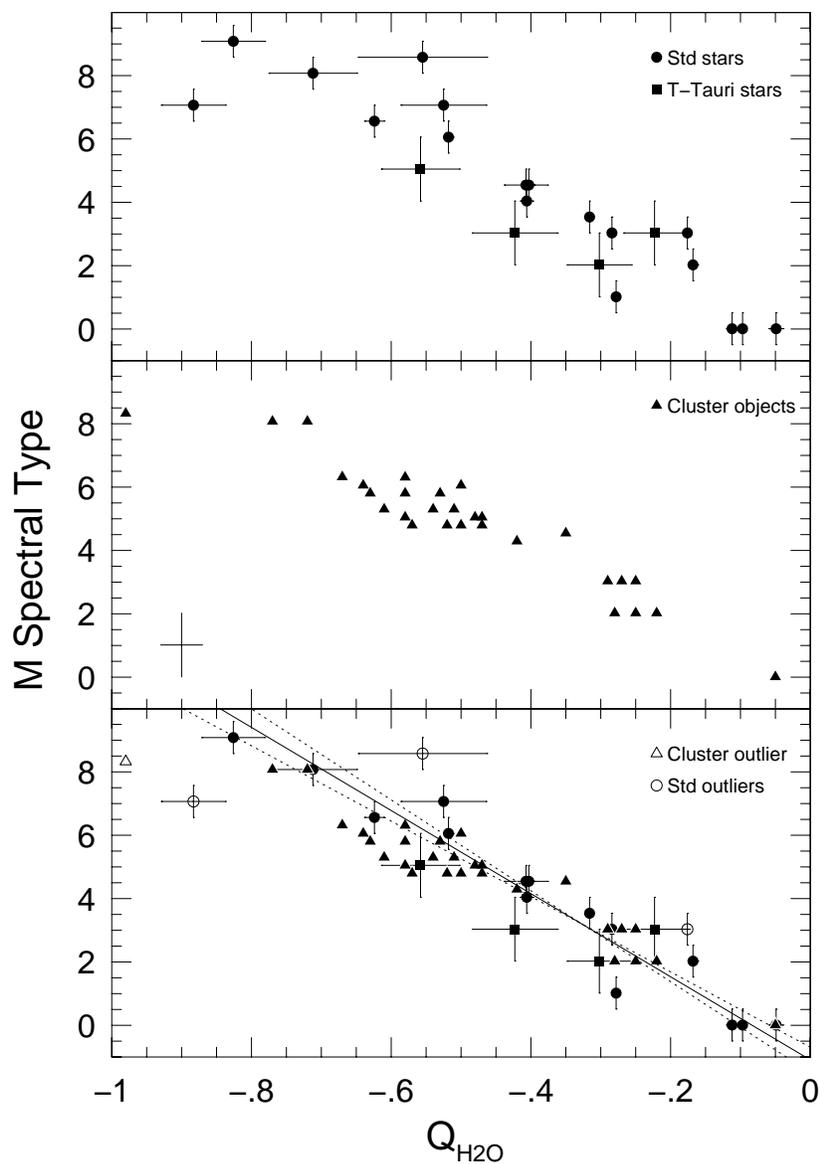}
{6.3in}{0}{65}{65}{-210}{-30}
\caption{The relationship between spectral type and $Q_{\rm H2O}$ for 
M dwarf standard stars and weak T Tauri stars (top panel) and for 
the IC348a stars in our sample with known optical spectral types 
(middle panel).  The standard and weak T Tauri stars have 1--$\sigma$ 
errors on $Q_{\rm H2O}$ as indicated; the spectral type errors shown 
are values from the literature.  In the middle panel, the typical 
spectral type errors (from the literature) and 
1--$\sigma$ errors on $Q_{\rm H2O}$  
are indicated in the lower left corner.
The bottom panel shows linear fits to the combined standard star and 
IC348a samples.  Open symbols indicate outliers.  Fits were performed 
in both senses (dotted lines) and the bisector (solid line) 
adopted as the calibrated relation.
}
\end{figure}

To explore the utility of the water index as an indicator of spectral 
type, we examined the relation between $Q_{\rm H2O}$ and spectral type for both 
the standard stars and a subset of IC348 stars that have optically determined
spectral types from \cite{luh98} and \cite{luh99}.  For the standard stars, 
we adopted spectral types from the literature that are derived consistently 
from the classification scheme of \cite{kir95}.  As shown in the top panel 
of Figure~\ref{qfit}, $Q_{\rm H2O}$ is strongly correlated and varies 
rapidly with spectral type among the standard stars, confirming the 
expected sensitivity of the water band strength to stellar effective 
temperature.  As is evident, there is real scatter 
among the standard stars that cannot be explained by errors in $Q_{\rm H2O}$ 
and spectral type.  The scatter may reflect the inherent diversity
in the standard star sample, a property that is evident from their $JHK$ 
colors.  
The spread in broad band color for a given spectral type is
usually interpreted as the result of varying metallicity
(e.g., Fig.~1 from \cite{leg96}).

To compare these results with those for a population that has a more 
homogeneous metallicity distribution and the same mean metallicity and gravity 
to the IC348 sample, we also examined the $Q_{\rm H2O}$ vs. spectral type 
relation for the subset of IC348 stars that have optical spectral types 
determined by \cite{luh98} and \cite{luh99} (middle panel of Fig.~\ref{qfit}).  
Although \cite{luh98} found no systematic difference between their IR and 
optical spectral types, there is significant dispersion between the two 
systems (their IR spectral types differ from the optical spectral types 
by as much as 3 subclasses). 
We find that the water band strengths are better correlated with the optical 
spectral types, with a smaller dispersion, than the IR spectral types,
suggesting that their optical spectral types are more precise.

With the use of optical spectral types, we were also able to compare
directly the results for the dwarf standards and the IC348 population,  
since both sets of objects are classified on the same system.
The two samples exhibit a similar relation between spectral type and $Q_{\rm H2O}$ 
despite the difference in gravity between the two samples, with some evidence
for a shallower slope for the pre-main sequence stars compared to the dwarfs.  
However, with the present data alone, we cannot claim such a difference 
with much certainty because the sample sizes are not large enough, the 
IC348 stars are not distributed evenly enough in spectral type, and there 
could be small systematic differences in the spectral typing of the IC348 
and standard stars.
The possibility of a difference between the two relations could be explored 
with more extensive optical spectral typing of the IC348 population.  

The horizontal and vertical error bars in the lower left corner of the
middle panel of Fig.~\ref{qfit} 
represent the typical errors in $Q_{\rm H2O}$ and spectral type for 
the cluster stars.  Some of the scatter may arise from infrared excesses 
(which would uniquely affect the young star sample, compared to the 
standard star sample), although this effect is expected to be limited 
given the relatively small fraction of cluster sources that have IR excesses.  
For example, based on their $JHK$ photometry, Lada \& Lada (1995) 
determined that $<$12\% of sources brighter than $K=14$ in IC348a 
have substantial IR excesses. The \cite{luh98} study 
spectroscopically inferred $K$ continuum excesses in a similar fraction 
(15\%) of sources in the subcluster. 

To examine the possible impact of IR excess on our derived $Q_{\rm H2O}$ values, 
we considered 
excesses of the form $\Delta F_\nu(\lambda) \propto \lambda^{-\beta}$ 
and explored 
the effect of the excess on the $Q_{\rm H2O}$ values for two of our standards,  
the M3 dwarf Gl388, and the M6 dwarf Gl406.
Since classical T Tauri stars have excesses at $K$ of $r_K\sim 0.6$ 
(\cite{mey97}; where $r_K$ is the ratio of the excess emission to 
the stellar flux), 
the IC348 sources, being more evolved, are likely to 
have much weaker excesses, typically $r_K < 0.2$. 
With a spectral index of $\beta = 1/3,$ appropriate for both disks 
undergoing active accretion and those experiencing passive reprocessing   
of stellar radiation, an IR excess produces an increase in $Q_{\rm H2O}$.  
Since the spectral slope is shallow and the maximum excess is small, 
only modest excursions are possible.  For example, 
the $Q_{\rm H2O}$ index for Gl388 varies from its observed value, -0.28, at 0\% 
excess to -0.24 at 20\% excess in F215N.  Over the same range of 
0 to 20\% excess in F215N, 
the $Q_{\rm H2O}$ index for Gl406 ranges from -0.52 to -0.43.  
This range of variation is sufficiently large that IR excess could 
account for most of the scatter of IC348 stars away from the mean 
trend to larger values of $Q_{\rm H2O}$.  
Explaining the scatter to smaller values of $Q_{\rm H2O}$ as the 
result of IR excesses requires more extreme values of $\beta$.  
For Gl388, values of $\beta < -3$ 
are needed to decrease $Q_{\rm H2O}$ from its value at 0\% excess.
Such extreme spectral indices are unlikely as they would produce 
unusual broad band colors. 
For these reasons, it appears unlikely that IR excess is responsible 
for the majority of scatter about the mean relation between $Q_{\rm H2O}$ 
and spectral type. 
Other processes are implied, possibly including those that produce 
true differences in stellar water band strengths among stars with 
equivalent $I$-band spectral types. 

Since we were not able to distinguish a systematic difference 
between the mean trends for the standard star sample and the IC348 
sample, we used the combined samples to calibrate the relation between 
$Q_{\rm H2O}$ and spectral type (lower panel of Fig.~\ref{qfit}).
In order to use the error information in both $Q_{\rm H2O}$ and spectral type, 
we performed a linear fit in both senses 
(i.e., spectral type vs. $Q_{\rm H2O}$ and 
$Q_{\rm H2O}$ vs. spectral type; dotted lines in Fig.~\ref{qfit}) and 
used the bisector of the two fits as the calibration relation 
(solid line in Fig.~\ref{qfit}).  
Due to the non-uniform distribution of stars
along the fit, the slope of the fit is sensitive to the
inclusion or exclusion of stars near the sigma-clipping limit and
at the extremes of either $Q_{\rm H2O}$ or spectral type.  
Doing a fit in both senses, and including the error information in 
both quantities, allowed us to better identify and exclude outliers. 
In the lower panel of Fig.~\ref{qfit}, 
solid symbols indicate the stars that were included in the fit while 
open symbols indicate excluded stars.

The equation of the bisector, the relation 
we subsequently used to estimate spectral class for the entire cluster 
sample, is:
\begin{equation}
{\rm M~subtype}=-1.09(\pm 0.39) - 13.01(\pm 0.50)\times Q_{\rm H2O}.
\label{eq:SpTQ}
\end{equation}
For a typical value of $Q_{\rm H2O}$, the formal
spectral type uncertainty in the fit is $\pm 0.46$, while the scatter
about the fit is $0.81$, just a little 
under one subtype.  It is noteworthy that the discrete nature of spectral 
type versus the continuous nature of $Q_{\rm H2O}$ is responsible for a 
mean scatter of $0.77$ in $Q_{\rm H2O}$ in each subtype bin, 
which is a significant contribution to the total scatter.  

Finally, we note that stars earlier than M2 have less certain spectral 
types due to the combination of the inherent scatter in the 
$Q_{\rm H2O}$ vs. spectral type relation and  
the decreasing sensitivity of the $1.9~\mu$m H$_2$O absorption band 
to spectral type as the K spectral types are approached.  
As a result, stars with spectral types of K and earlier
can be misclassified by our method as later-type objects. 
For example, 
a comparison of the spectral types obtained by LRLL and \cite{luh99}  
with those obtained by our method shows that 
stars earlier than $\sim$K5 are classified by us as late K or M0 stars
and late-K stars are classified as late-K and M0-M1 stars.

\section{Extinction}

\begin{figure}
\figurenum{8}
\label{red}
\plotfiddle{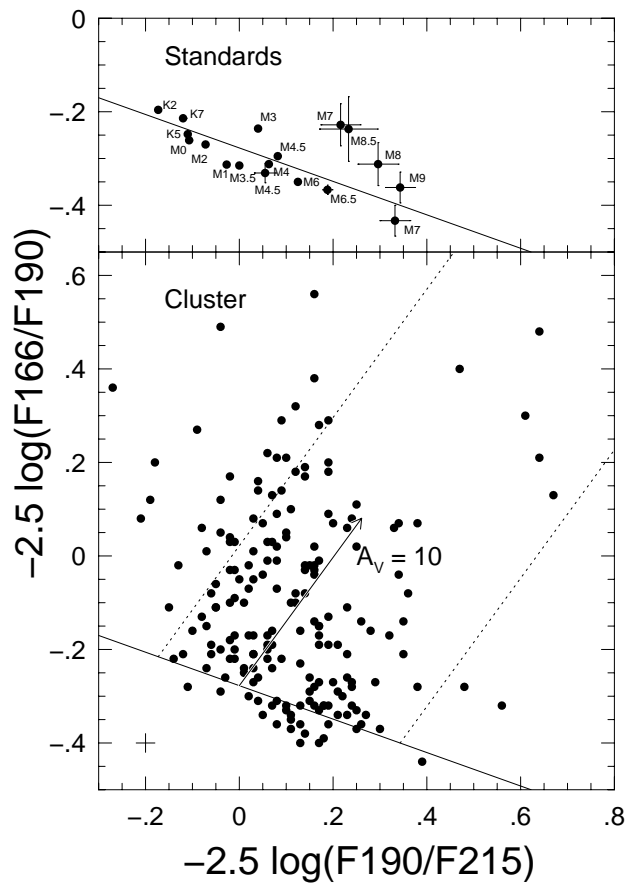}
{6.5in}{0}{65}{65}{-210}{40}
\caption{Determination of extinction from the narrow-band colors.
The top panel shows the zero-reddening line determined from a 
least-squares fit to the dwarf standards.  Objects without apparent 
error bars have photometric errors smaller than 
the point size.  The extinction for the subcluster stars 
is determined by dereddening the stars to the zero-reddening line 
(bottom panel).  Typical errors in the colors of the subcluster stars, 
representative of all but the faintest cluster stars, are shown in the 
lower left corner.
}
\end{figure}

Although the stellar spectral typing could be carried out 
without determining the reddening to each object, 
extinction corrections are required in order to investigate the 
masses and ages of cluster objects.
We estimated the extinction toward each star by 
dereddening the observed F166/F190 and F190/F215 colors to 
a fiducial zero-reddening line in the color-color plane.  
Since extinction estimates for the dwarf standard stars were not 
available in the literature, we adopted the usual assumption that 
they suffer zero extinction. 
Figure~\ref{red} diagrams the process.  First, we fit a line to the 
positions of the standard stars in the color-color plane (top panel), 
which is defined to be a locus of zero reddening.
The WTTS were excluded from the fit.  Gl569A was regarded as an outlier 
and also excluded from the fit. 
The resulting linear relation is:
\[
-2.5\log{\left(\frac{\rm F166}{\rm F190}\right)}=-0.277(\pm0.009)
-0.358(\pm0.083)\times- 2.5\log{\left(\frac{\rm F190}{\rm F215}\right)}
\]
with a mean deviation about the fit of 1--$\sigma = 0.036$.  
The extinction toward each star in the cluster fields was determined 
from the shift in each color required to deredden the star to the 
zero-reddening line.  
The resulting extinction estimates and errors are given 
in column 11 of Table~\ref{astrometry}.  Note that the reddening vector 
(shown for $A_V = 10$ in the bottom panel of Fig.~\ref{red}),  
is nearly perpendicular to the standard star locus 
in the color-color plane.  
Consequently, 
reddening and spectral type are readily separable 
with moderate signal-to-noise photometry even given modest uncertainties 
in the slope of the reddening vector. 

The subset of our standards used for the reddening calibration span
the spectral class range K2V to M9V.  This range is indicated 
by the dotted lines in the lower panel of Fig.~\ref{red}. 
The few stars in the field with spectral types outside this range 
have
extinction estimates based on the extrapolation of the fiducial line.
As we show in Section 8, most of the early type stars are likely 
background objects.  
Finally, while the formal uncertainty in the fit of the fiducial line
to the standards is small, 0.04 magnitudes, the scatter about the line
for the latest standards is significantly larger than the scatter for
the earlier standards (top panel of Fig.~\ref{red}).  Part of this 
scatter is due to the larger photometric errors; the late
type standards are also the dimmest.  However, four of the five late type
standards fall above the fiducial line.  In order from upper left to 
lower right these standards are LHS3003(M7V), Gl569B(M8.5V), VB10(M8V), 
LHS2924(M9V), and VB8(M7V).  With the exception of VB8, these stars
are aligned in the expected order in both colors but seem to be systematically
shifted about 0.1 magnitudes to the red in $-2.5\log{({\rm F166/F190})}$.  
While we cannot exclude the possibility that the relationship is non-linear
for dwarfs later than M6, some of the scatter about the fit may be due to 
inherent variation in the photometric properties of the standard stars. 

\begin{figure}
\figurenum{9}
\label{redcomp}
\plotfiddle{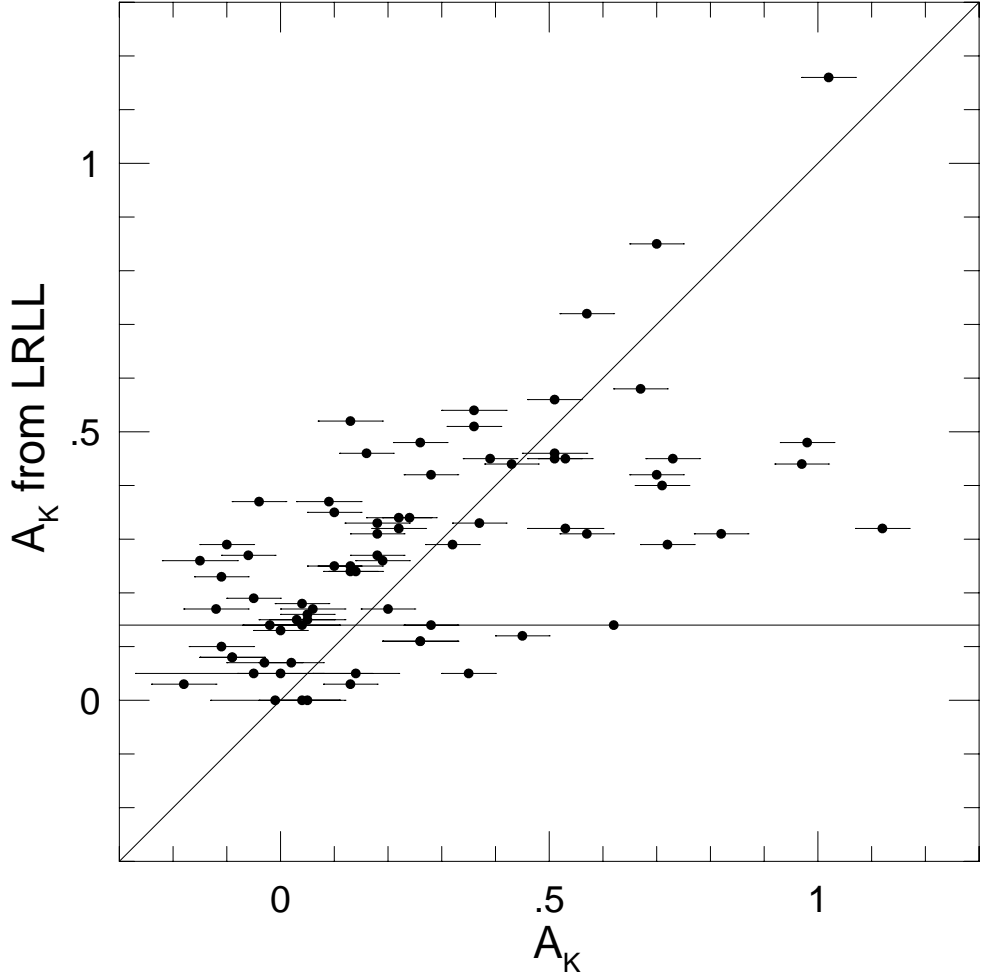}
{6.0in}{0}{75}{75}{-235}{-40}
\caption{Comparison of our extinction values with those of \cite{luh98} 
for the M dwarfs common to both samples.  The error bars indicate the 
formal uncertainty in our $A_K$ estimates.  The $A_J$ values from 
\cite{luh98} were converted assuming standard interstellar reddening 
$A_K=0.37 A_J$.
The diagonal line is unity.}  
\end{figure}

We can compare our extinction estimates to those of \cite{luh98} for the 
M dwarfs common to both samples.  In Figure~\ref{redcomp}, the horizontal 
error bars indicate the formal (1--$\sigma$) uncertainty in our 
$A_K$ estimate (typically $<0.1$ mag). 
\cite{luh98} used various extinction estimators, citing their internal 
errors rather than values for individual stars.  Their errors in $A_K$ 
range from 0.07 to 0.19 mag for the stars shown with a nearly equal 
systematic uncertainty in the zero point.  
For the M-dwarfs common to both samples, the mean difference in 
$A_K$, in the sense $\langle{\rm Ours - LRLL}\rangle = -0.01 \pm 0.03$ 
with a scatter about the mean of 0.24 magnitudes.  Considering the 
uncertainties, the agreement is good.

\begin{figure}
\figurenum{10}
\label{redhisto}
\plotfiddle{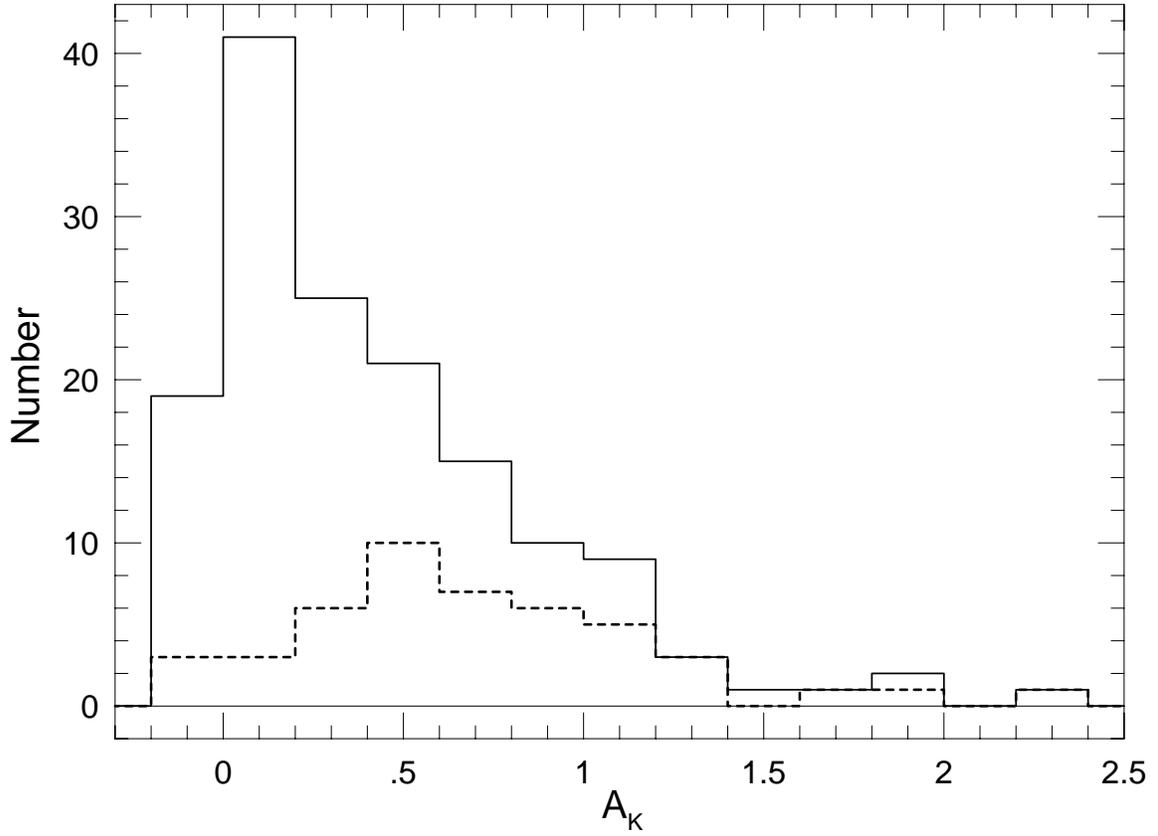}
{5.5in}{0}{65}{65}{-260}{20}
\caption{The distribution of $A_K$ values for the IC348a stars 
that fall between the dotted lines in Fig.~\ref{red} (solid line histogram)
and for the subset of objects identified as the background population 
(dashed-line histogram).  
}
\end{figure}

The resulting $A_K$ distribution 
(Fig.~\ref{redhisto}; solid-line histogram), 
has a pronounced tail to large values of $A_K$ and a peak at $A_K = 0.1$.  
The extinction distribution for the subset of objects 
identified as the background population 
(as determined in section 8; dashed-line histogram) is also shown.  
Note that our extinction estimates include a few negative values 
(Figs.~\ref{red} and~\ref{redcomp}).  
While these values might suggest that our fiducial line needs to be lowered
to bluer colors, that would imply a bias toward larger extinctions given the 
distribution of standard stars in the color-color plane.  Therefore, we 
retain our original fit and, for all subsequent analysis, stars with negative 
extinction estimates are assigned an extinction of 0.0 with an error equal 
to the greater of the absolute value of the original extinction estimate or 
the formal uncertainty in the estimate.

With this revision, the mean extinction is $\langle A_K \rangle = 0.44$ with an
error in the mean of 0.04 and a median of $(A_K)_{\frac{1}{2}} = 0.31$.
Our adjustment of the negative values impacts negligibly on the statistics. 
(If the negative extinction values were retained, the mean would be 
$\langle A_K \rangle = 0.43$ with the error and median unchanged.)  When our 
sample is restricted to those stars in common with \cite{luh98}, we
find approximately the same mean reddening ($\langle A_K \rangle = 0.30$)
that they quote for their sample ($\langle A_K \rangle = 0.34$).  
The larger mean reddening in the present study 
indicates that, on average, we have sampled 
a more extincted population of the cluster than
has been investigated previously.
Using the position of the main sequence at the distance of the cluster
(see section 8)
to divide the sample into cluster and background objects, we find that 
the cluster objects have
$\langle A_K \rangle = 0.31 \pm 0.04$ with a scatter about the mean of
0.36.  The background stars, which include most of the stars in the
extended high extinction tail, have $\langle A_K \rangle = 0.71 \pm 0.07$,
with a scatter about the mean of 0.49.

\begin{figure}
\figurenum{11}
\label{redmap}
\plotfiddle{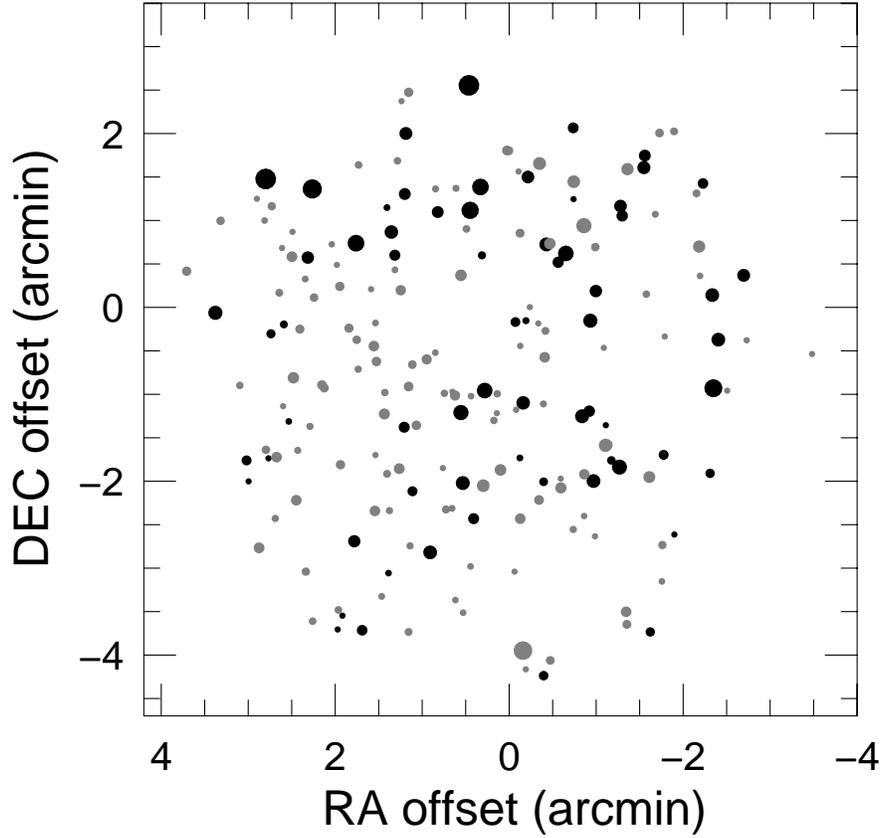}
{4.5in}{0}{80}{80}{-325}{-30}
\caption{The spatial distribution of extinction and sample membership
for all of the stars in our sample.  
The point size is scaled to the estimated extinction to each object 
(larger points corresponding to larger reddening),  
which ranges from $A_K = 0.00$ to 2.33.  
The gray points represent objects in common with previous studies 
(\cite{luh98}; \cite{her98}; \cite{luh99}; see Table~\ref{astrometry}).  
The black points represent objects without previous reddening estimates.  
}
\end{figure}

The more heavily reddened stars in our sample are spatially intermixed with 
stars experiencing lower extinction. 
Figure~\ref{redmap} shows the same area plotted in Fig.~\ref{finder}.  
The gray symbols denote stars in our sample that were 
observed by other investigators (\cite{luh98}; \cite{her98}; Luhman 1999),  
whereas the black symbols denote stars that were not 
observed by these investigators. 
The point size is scaled to our estimate of the extinction 
to the object (larger points corresponding to larger reddening),  
which ranges from $A_K = 0.0$ to $A_K = 2.33$.  
The higher average extinction among the black points is apparent.
The extinction distribution is characterized by an overall gradient 
from NE (larger values) to SW (smaller) with significant small scale 
variation. 
Given the broad extinction distributions for both cluster and 
background objects, 
and the patchy distribution of extinction on the sky, 
it is evident that cluster membership cannot be 
determined on the basis of extinction alone.  Membership based on 
extinction would erroneously assign low extinction background members 
to the cluster and highly extincted cluster members to the background.

\section{Observational HR Diagram }

\begin{figure}
\figurenum{12}
\label{hrd}
\plotfiddle{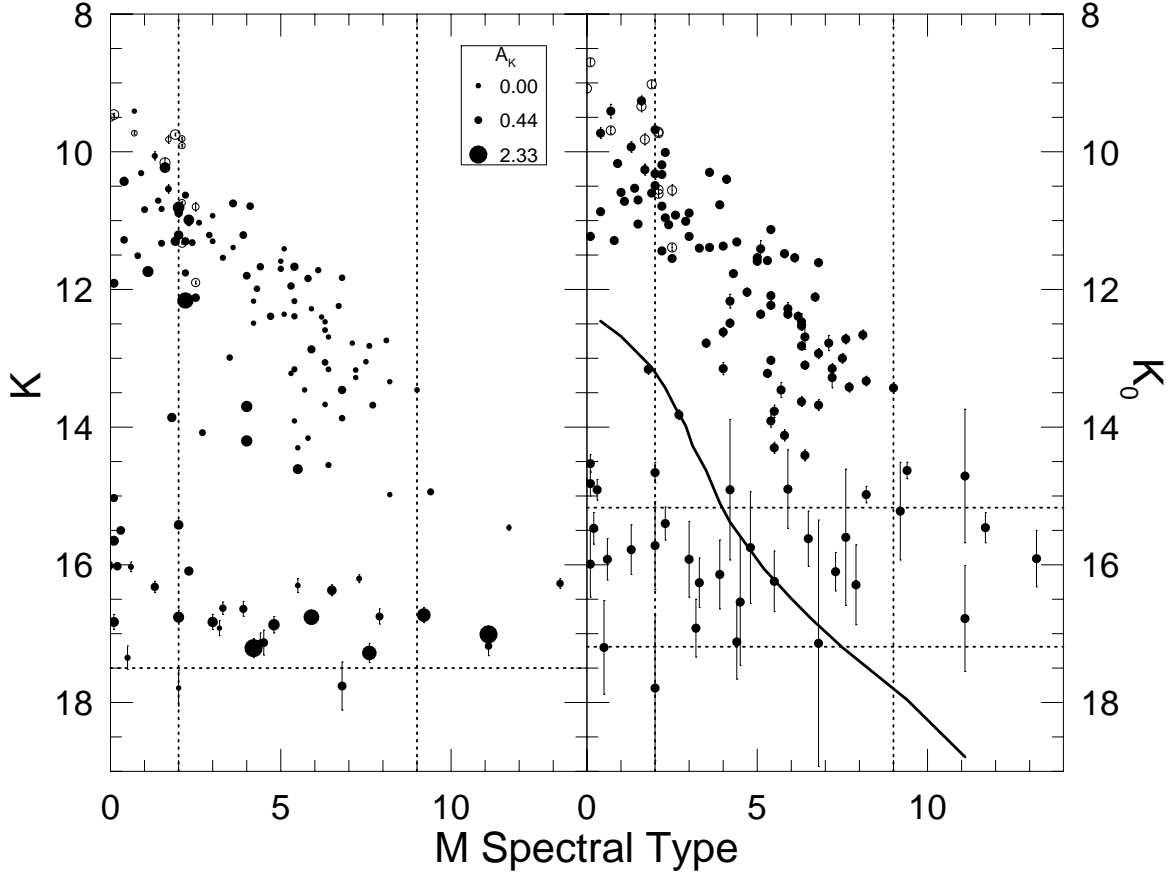}
{5.0in}{0}{70}{70}{-285}{-10}
\caption{Observational HR diagrams of stars in the IC348a region with 
spectral types M0 and later.  The left panel plots observed $K$ magnitude
against $Q_{\rm H2O}$ spectral type.  The symbol size indicates the 
estimated reddening toward each star (see key).  The vertical 
error bars indicate the photometric uncertainty.  
Stars without apparent vertical 
error bars have photometric uncertainty smaller than the point size.
The vertical dotted lines indicate the range over which the $Q_{\rm H2O}$ 
spectral types are well calibrated.  
The right panel plots dereddened $K$ magnitude against $Q_{\rm H2O}$ 
spectral type.  The vertical error bars include both photometric 
and extinction uncertainties.  
The solid curve represents the main sequence at the 
distance of the cluster, $(m-M)_0 = 7.4$.  
In both panels, horizontal dotted lines indicate our photometric 
completeness limit (see text), and  
stars plotted as open symbols indicate stars whose 
$Q_{\rm H2O}$ spectral types were subsequently replaced 
with optical spectral types from \cite{luh98} or \cite{luh99} (see text).} 

\end{figure}

With the spectral types determined in section 5 and the extinction
extimates from section 6, we can construct an observational HR diagram 
of the cluster fields.  In Figure~\ref{hrd}, the vertical 
axes are apparent $K$ magnitude (left panel) and 
dereddened $K$ magnitude, $K_0$ (right panel).  
For comparison, the solid curve in the right panel is the 
fiducial main sequence at the distance of the cluster (see section 8.2).
Examination of both panels reveals a well defined 
cluster sequence at $K\leq 14$.  This locus is marginally tighter after 
being dereddened which supports the accuracy of our reddening estimates.  

Spectral type errors are not shown, both to limit confusion
and because some stars have systematic as well as random error.
For example, although the typical random error is $\pm 1$
spectral subtype, stars earlier than M2 have systematically 
later $Q_{\rm H2O}$ spectral types than optical spectral types (section 5).
Given the possible inaccuracy of our spectral typing scheme at 
spectral types earlier than M2, we adopted the optical 
spectral types of \cite{luh99} or \cite{luh98} for these objects where 
available.  The original $Q_{\rm H2O}$ spectral types of these stars 
are shown as open circles in Figure~\ref{hrd}.  
When optical spectral types of these stars are adopted instead (see 
subsequent figures), the photometric width of the distribution at M2 
and earlier is reduced. 
In general, the random error in spectral type increases with increasing 
magnitude (see column~13 of Table~\ref{astrometry}).  All stars with 
$K < 15.5$ have spectral type errors $\leq 1$ subtype.  
Since our spectral type errors grow rapidly below $K=16$, with 
stars fainter than $K=16.5$ having spectral type errors 
$\gtrsim 2.5$ subtypes, we identify $K=16.5$ as our 
effective magnitude limit for accurate spectral typing.  

While some objects have spectral types as late as ``M13'', this should 
be interpreted simply as an indication of strong water absorption rather 
than an advocacy of M spectral types beyond M9. 
The existence of objects with stronger water absorption than that of 
M9 dwarfs is in general agreement with the predictions of atmospheric models  
(e.g., the Ames-Dusty and Ames-MT-Dusty models of \cite{all00}). 
These suggest that even in the presence of dust, 
the $1.9~\mu$m H$_2$O absorption band continues to increase in strength 
down to $\sim 2000$ K at pre-main sequence gravities.  
In the Ames-Dusty models, $Q_{\rm H2O}$ increases in strength by 45\% 
between 2450K (equivalent to M8 in the dwarf temperature scale; 
see section 8.2) and 2000K.  
The $Q_{\rm H2O}$ vs.\ spectral type relation in 
eq.~\ref{eq:SpTQ} implies that $Q_{\rm H2O}$ is 54\% stronger at M13 than 
at M8, in general agreement with the predictions.  

The dearth of stars at $K\approx 15.5$ in the $K$ luminosity function 
is also evident in the left panel of Fig.~\ref{hrd}.  
Part of the deficit is due to the higher average reddening of the 
background stars.
Stars with $K > 15.5$ have an average extinction greater
than stars with $K < 15.5$ and when they are dereddened, they fill in 
the deficit somewhat.
Our photometric completeness limit of $K=17.5$
is shown in the left panel as a horizontal dotted line.  
To quantify our detection limit as a function of extinction,  
we also show the completeness limit dereddened by $A_K=0.31$, the mean 
extinction among the cluster stars  
(lower horizontal dotted line in the right panel of Fig.~\ref{hrd}) 
and by $A_K=2.33$, the greatest
extinction detected in the cluster fields 
(upper horizontal dotted line). 
Both limits, $K=17.19$ and $K=15.17$,
are considerably dimmer than the typical cluster M star.

These results imply that we have fully sampled the cluster population 
over a significant range in extinction.  The extinction range that we 
probe is, of course, a function of spectral type.  As examples, 
of the two cluster stars in the tail of the reddening distribution 
shown in Fig.~\ref{redhisto}, one is an M2 star with $K = 12.16$ 
($A_K=1.97$) and the other is an M9 star with $K = 16.73$ ($A_K=1.52$).  
We would have been able to detect and spectral type the first
star through another $\sim 4.4$ mag of extinction (to $A_K \approx 6.4$).
The second star, observed through almost 5 times the average cluster 
extinction, is close to our spectral typing limit. 

\section{Comparison with Evolutionary Tracks}

\subsection{Evolutionary Models}

Evolutionary models for low mass objects 
have developed greatly in recent years, with several 
different models now available over a large range in mass.  
\cite{dm97}
have recently updated their pre-main sequence calculations,  
retaining the use of the Full Spectrum Turbulence model of 
\cite{can91}
and making improvements in opacities and the equation of state.
For the purpose of this paper, we use their 1998 models 
\footnote{These models are available at:
http://www.mporzio.astro.it/~dantona} 
(hereinafter DM98)
which cover the mass range $0.017-0.3 M_\odot$
and include further improvements, 
e.g., in the treatment of deuterium burning,  
that affect the very low mass tracks.  

Other groups (e.g., \cite{bar98}; Burrows et al.\ 1997) 
have also presented new evolutionary models that include improvements 
in the treatment of the stellar interior 
and use non-gray atmospheres as an outer boundary condition.  
The corrections associated with the latter are particularly 
significant at low masses since the presence of molecules in low 
temperature atmospheres results in spectra that are significantly 
non-blackbody.  Models by Baraffe et al.\ (1998; hereinafter B98) 
explore the mass range $\sim 0.025-1.0 M_\odot$ 
using the Allard et al.\ (1997) NextGen synthetic atmospheres.  
Although there are known inconsistencies in the NextGen models 
(e.g., they overpredict the strength of the IR water bands; 
TiO opacities are suspected to be incomplete; 
grain formation is not included), the B98 models nevertheless 
reproduce well the main sequence 
properties of low metallicity populations, e.g., the 
optical color-magnitude diagram of globular clusters and 
halo field subdwarfs.  There is also good agreement with the 
optical and IR properties of nearby disk populations, although 
some discrepancies remain at low masses ($< 0.15 M_\odot$).

Non-gray models have been developed independently by 
\cite{bur97} who focus on the properties of 
objects at lower mass ($0.3-70 M_J$, where $M_J$ is the mass of Jupiter).
The Burrows et al. evolutionary tracks differ qualitatively 
from those of B98 in the upper mass range, 
but are more qualititatively similar at masses $\lesssim 60 M_J.$ 
The qualitative difference between these models, which appear to have 
similar input physics, may indicate the current level of uncertainty 
in the evolutionary tracks at low masses.
Quantitatively, an effective temperature of 3340 K and 
luminosity of $0.076 L_\odot$ 
corresponds to a mass and age of $0.090 M_\odot$ and 1.8 Myr 
with the Burrows et al. tracks 
and $0.3 M_\odot$ and 8 Myr with the B98 tracks. 
The tracks agree better in mass
in the lower mass range: at 2890 K and $0.022 L_\odot$, 
Burrows et al. predict $0.05 M_\odot$ at 1.2 Myr, 
and the B98 tracks predict $0.06 M_\odot$ at 3.2 Myr.

\subsection{Interloper Population}
As reviewed by Herbig (1998), 
the distance to IC348 has been previously estimated on the basis of 
both nearby stars in the Per OB2 association 
and stars in the IC348 cluster itself.
For the purpose of comparing our results with evolutionary tracks, 
we adopt a distance to IC348 of $d = 300$ pc, $(m-M)_0 = 7.4$.
This value is in good agreement with current estimates of the 
distances to the Per OB2 cluster 
(318$\pm 27$ pc; de Zeeuw et al.\ 1999) 
and to IC348 itself (261$\pm 25$ pc; Scholz et al.\ 1999)
inferred from Hipparcos data.
The adopted distance is also in agreement with the value adopted by both
Herbig (1998) and \cite{luh98} and thereby allows ready
comparison of our results with those obtained in previous studies.

To delineate the background population,
the position of the main sequence at the cluster distance is
indicated by the solid curve in the right panel of Fig.~\ref{hrd},
where we have used the 12 Gyr isochrone from the B98 evolutionary tracks 
and a temperature scale that places the isochrone in good 
agreement with the main sequence locus 
of nearby field stars (e.g., \cite{kir94}).
The temperature scale used, 
\[
{\rm M~subtype} = (4000 - T_{\rm eff})/180,
\]
is generally consistent with the Leggett et al.\ (1996) 
dwarf temperature scale.

The magnitude and spectral type distributions of the background
population, located to the lower left of the main sequence,
are in very good agreement with the total interloper population predicted by
models of the point source infrared sky (\cite{wai92}; Cohen 1994)
at the Galactic latitude and longitude of IC348. 
Table~\ref{cohenbck} compares the observed and model counts as a function 
of $K$ magnitude and spectral type. 
To $K_0=17$, significant departures between the model and observed counts
are apparent only for spectral types earlier than M3 at $K_0>16$.
Given the large spread in the reddening distribution of the background
population (to $A_K>2$; Fig.~\ref{redhisto}),  
this discrepancy in the counts probably arises from photometric
incompleteness below $K=17.5$.
This result (the good agreement between the model prediction for the 
total interloper population and the observed background population),
implies a negligible foreground contamination (at most $1-2$ stars) 
of the cluster population at late spectral types.
The large reddening of many of the faint late-type stars also statistically
argues against a foreground origin for these objects.
Note, however, that the errors on some of the fainter objects identified
as older cluster members (e.g., objects in the range $K_0=15.5-16.5$,
M6$-$M8) allow for the possibility that they are
background objects even if they are not predicted to be so by the
Galactic structure model.

\subsection{Temperature Scale and Bolometric Correction}

A generic difficulty in comparing measured stellar fluxes and
spectral types with evolutionary tracks is the need to adopt
relations between spectral type, effective temperature, and
bolometric correction.  
In principle, such relations could be avoided by using synthetic
spectra from model atmospheres to go directly from observed spectra
and colors to temperature and gravity, and hence to mass and age 
using the theoretical evolutionary tracks.
For example, we might hope to compare directly the water band
strengths of the Allard \& Hauschildt atmospheres used in the B98
models with the water band strengths that we measured.
However, since there remain significant quantitative differences 
between the predicted and observed water band strengths of M stars 
(e.g., the models consistently overpredict water band strengths; 
see also \cite{tie00}), 
this approach cannot be used in the present case. 
In other words, although current synthetic atmospheres may be 
sufficiently accurate for the purpose of evolutionary calculations and 
the prediction of broad band colors, they are insufficiently 
accurate as templates for spectral typing.  
Hence, we adopted the less direct method of first calibrating our
water index versus spectra type (section 5), and then selecting
an appropriate spectral type to temperature conversion.

Ideally, we would want to use a relation between spectral type and 
effective temperature that is 
appropriate to the gravity and metallicity of the IC348 population.
Unfortunately, an empirical calibration of spectral type and effective temperature
appropriate for pre-main-sequence conditions has yet to be made.
In the meantime, since pre-main-sequence gravities are similar to dwarf
gravities, temperature scales close to the dwarf scale
(e.g., \cite{leg96}) are often used in the
study of young populations (e.g., LRLL; \cite{wil99}).
Because the temperature scale may differ from that of dwarfs 
at PMS gravities, other choices have also been investigated,
including temperature scales intermediate between those of dwarfs
and giants (e.g., \cite{whi99}; Luhman 1999).  

\begin{figure}
\figurenum{13}
\label{ObsHRD}
\plotfiddle{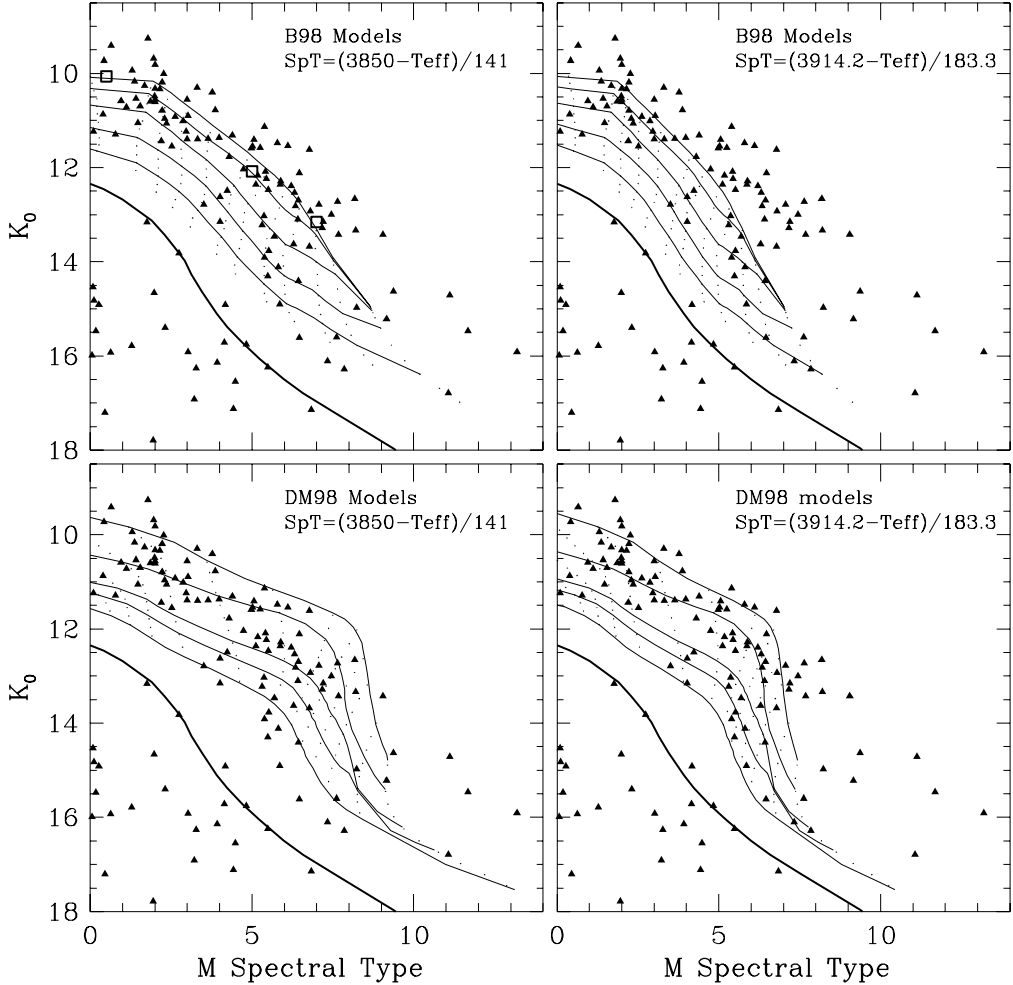}
{5.0in}{0}{75}{75}{-225}{-100}
\caption{Observational HR diagrams of stars in the IC348a fields (triangles) 
compared with 4 combinations of evolutionary models and temperature 
scales. 
{\it Upper left}: the B98 models and the 
\cite{luh99} intermediate temperature scale.  
Isochrones (light solid curves) are shown for 2, 3, 5, 10 and 20 Myr.
Mass tracks (dotted curves) are shown for 
0.025, 0.040, 0.055, 0.075, 
0.1, 0.15, 0.2, 0.3, 0.4, 
0.5, 0.6, and 0.7~$M_\odot$; 
the $\alpha$=1.9 track is used for $0.7 M_\odot$. 
Three of the components of GG Tau are also shown (open squares). 
{\it Upper right}: the B98 models and the dwarf temperature scale.  
{\it Lower left}: the DM98 models and the \cite{luh99} intermediate 
temperature scale.  
Isochrones (light solid curves) are shown for 0.3, 1, 3, 5, and 10 Myr.
Mass tracks (dotted curves) are shown for 
0.017, 0.025, 0.04, 0.055, 
0.075, 0.1, 0.14, 0.2, 0.3, 
0.4, 0.5, and 0.6~$M_\odot$;
the DM97 tracks are used for $M\ge 0.4 M_\odot$. 
{\it Lower right:} the DM98 models and the dwarf temperature scale.  
In each panel, the main sequence is represented by the heavy solid line.  
}
\end{figure}

The validity of the various evolutionary tracks 
can be evaluated by a number of criteria including
whether stellar masses predicted by evolutionary tracks
agree with dynamical estimates, 
and whether populations believed to be coeval appear so when
compared with evolutionary tracks (e.g., Stauffer et al.\ 1995).
Dynamical mass constraints are becoming available in the
$1 M_\odot$ range (see, e.g., \cite{mat00}) but are thus
far unavailable at the masses of interest in the present study.
In contrast, coeval population constraints are more readily 
available at these lower masses. 
For example, in the GG Tau hierarchical quadruple system
(\cite{whi99}),
the four components of the system, arguably coeval, span a wide range
in spectral type (K7 to M7; open squares in Fig.~\ref{ObsHRD}, upper left), 
thereby
outlining, in rough form, an isochrone spanning a large mass range.
When plotted at a common distance, the IC348 cluster locus identified
in the present study overlaps the locus defined by
the GG Tau components over the same range of spectral types (Fig.~\ref{ObsHRD}).
This both reinforces the validity of the GG Tau system as a coeval
population constraint and argues that the mean age of the IC348 cluster
is approximately independent of mass.
Similar results have been found previously at spectral types earlier
than M6 (Luhman 1999).  

The uncertainty in the pre-main-sequence temperature scale complicates
our understanding of the validity of the tracks. 
As discussed by Luhman (1999),
combinations of evolutionary tracks and temperature scales
that are consistent with a coeval nature for the GG Tau system
and the IC348 cluster locus include
(1) DM98 tracks and a dwarf temperature scale
(2) B98 tracks and an otherwise arbitrary temperature scale intermediate
between that of dwarfs and giants.
Our results are compared in Figure \ref{ObsHRD}
with these combinations of
temperature scales and tracks.
For comparison, the two alternative combinations of temperature scales
and tracks are shown.
In comparing the B98 models with the observations, we have used
the model $K$ magnitudes and a linear fit to either the
dwarf temperature scale
\begin{equation}
{\rm M\ subtype} = (3914 - T_{\rm eff})/183.3 
\label{eq:teLeg}
\end{equation}
or the Luhman (1999) intermediate temperature scale
\begin{equation}
{\rm M\ subtype} = (3850 - T_{\rm eff})/141.0 
\label{eq:teL99}
\end{equation}
In approximating the dwarf temperature scale, particular weight was given 
to the dwarf temperature determinations by \cite{tsu96}
who used the IR flux measurement technique.  As they show, this technique 
is relatively insensitive to the details of synthetic atmospheres (e.g., 
dust formation).  The fit thus obtained is in good agreement with the 
temperature determinations of Leggett et al.\ (1996) which are based on 
a comparison 
of synthetic atmospheres with measured IR colors and spectra.  
In comparing the DM98 models with the observations, we have used,
in addition to these temperature scales, a bolometric correction
$$ {\rm BC_K} = M_{\rm bol} - M_K = 4.19  - T_{\rm eff}/2240$$
that extrapolates the values obtained by Leggett et al. (1996)
and \cite{tin93} to low temperatures.

The combination of the B98 models and the Luhman intermediate 
temperature scale (eq.~\ref{eq:teL99}; Fig.~\ref{ObsHRD} upper left) 
implies that the mean age of the cluster is approximately independent of 
mass over the range $0.7-0.04 M_\odot$.
The comparison implies a mean age $\sim$3 Myr with a 
age spread from $< 1$ to $\sim 20$ Myr.  
The faint cluster population between spectral types M5 and M8
appears to constitute an old cluster population ($\sim 5$ to $>20$ Myr) 
with masses $0.13-0.05 M_\odot$. 
If the dwarf temperature scale 
(eq.~\ref{eq:teLeg}; Fig.~\ref{ObsHRD} upper right) is used 
instead, the cluster is, 
on average, significantly younger at late spectral types. 

The combination of the DM98 models and the dwarf 
temperature scale (eq.~\ref{eq:teLeg}; Fig.~\ref{ObsHRD} lower right) 
implies that the mean cluster age is approximately independent of 
mass at spectral types earlier than M7
but younger at late types.
The comparison implies a mean age $\sim 1$ Myr with a 
age spread from $< 1$ to $\sim 10$ Myr.  
With these models, the faint cluster population between 
spectral types M5 and M8 is spread over a larger range in mass
$0.16-0.025 M_\odot$. 
If the Luhman intermediate temperature scale 
(eq.~\ref{eq:teL99}; Fig.~\ref{ObsHRD} lower left) 
is used instead, the cluster is older at late types with a 
larger spread in age.
With all combinations of models and temperature scales, 
the brighter cluster population beyond M8 is systematically younger, 
$< 1$ Myr old.  
If this is an artifact, it may indicate the likely inadequacy 
of the assumed linear relation between effective temperature 
and spectral type over the entire range of spectral types in the sample. 
Deficiencies in the evolutionary tracks are another possibility. 

It is interesting to examine the motivation for the intermediate
temperature scale adopted by White et al. (1999) and Luhman (1999).  
These authors have argued that since the M giant temperature scale is
warmer than the dwarf scale, PMS stars, which are
intermediate in gravity, may be characterized by a
temperature scale intermediate between that of giants and dwarfs.
\cite{luh99} has further shown that the spectra of pre-main sequence 
stars in IC348 are better fit by an average of dwarf and giant 
spectra of the same spectral type.

There are several caveats to this argument.
Firstly, the giant temperature scale considered by Luhman (1999)
is derived from the direct measurement of stellar angular diameters
(e.g., \cite{per98}; \cite{ric98}; \cite{van99}),
whereas the dwarf temperature scale is typically determined with the 
use of model spectra (e.g., \cite{leg96}; \cite{jon94}; \cite{jon96}). 
The different methods by which the two temperature scales are derived
may introduce systematic differences that do not reflect a true
temperature difference.

Secondly, we can turn to synthetic atmospheres for insight into
the gravity-dependent behavior of the temperature scale.
In the current generation of the Allard \& Hauschildt atmospheres
(e.g., Ames-Dusty, Ames-MT-Dusty),
the 1.9$\mu$m water band strength is relatively insensitive to gravity
above 3000K ($\sim$M5 in the dwarf scale).
At effective temperatures below 3000K, dust formation is significant,
introducing added complexity to the gravity dependence of the
atmosphere in the 1.9$\mu$m region.
In this temperature range, the water index first increases in strength
($Q_{\rm H2O}$ decreases) at fixed temperature 
from $\log g \sim 3.5$ to $\log g \sim 5.0-5.5$ 
(due to increased water abundance) 
then decreases in strength with higher gravity
(due to increased dust formation and consequent backwarming and
dissociation of water).
The net result is a {\it cooler} temperature scale
for pre-main-sequence gravities below 3000K. 
For example, at $\sim 2700$ K pre-main-sequence objects ($\log g = 3.5-4.0$) are
$\sim 200$K cooler than dwarfs ($\log g = 5.0-5.5$) with an equivalent
water strength.  

On the basis of these models, there is little physical motivation
for an intermediate temperature scale beyond M4 for the interpretation
of water band strengths.
Of course, these considerations apply to the interpretation of
1.9$\mu$m water band strengths rather than the 6500-9000\AA\ region 
studied by \cite{luh99}.  
A detailed examination of current synthetic atmospheres for the latter  
spectral region may provide better motivation for a hotter temperature 
scale at lower gravities.

Note that the gravity dependence of $Q_{\rm H2O}$ in 
the synthetic atmospheres is modest over the range of gravities 
relevant to low mass pre-main sequence 
stars in the age range of the cluster (1--10 Myr).  
For example, in the B98 model, an $0.06 M_\odot$ object follows 
a vertical evolutionary track at $T_{\rm eff} \sim 2860$K with 
$\log g=3.6-4.2$ 
in the age interval 1--10 Myr which corresponds to a fractional change 
in $Q_{\rm H2O}$ of $\lesssim  15$\% or $\lesssim  1$ subtype, given the 
relation between $Q_{\rm H2O}$ and spectral type discussed in section 5.

In summary, while we can find little physical motivation for an 
intermediate
temperature scale with which to interpret our results, we interpret
the better fit to the IC348 cluster locus that we obtain with 
the combination of this temperature scale and the B98 models as an 
indication of the direction in which the evolutionary model 
calculations might themselves evolve in order to
better reproduce observations of young clusters.
With these caveats in mind, we discuss, in the next section, 
the cluster mass function  
implied by 2 combinations of tracks and temperature scales.
However, it is already clear that there will be 
reasonable uncertainty associated with such results.

\section{Discussion}

\subsection{Binarity}

The area and depth that we have covered at relatively high angular 
resolution, combined with our ability to discriminate cluster members 
from background objects, allows us to place some useful constraints 
on the binary star population of the cluster.
At the pixel scale of NIC3,  
pairs of stars with separations $\gtrsim 0.8\arcsec$ are easily 
identified over the entire magnitude range of our sample;  
for fainter primaries, companions could be similarly detected at smaller 
separations.
A significant obstacle to the detection of faint companions 
at separations $\lesssim 0.8\arcsec$ 
is the complex, extended structure in the NICMOS PSF 
which also makes it difficult to quantify our detection completeness. 
More refined techniques, such as PSF subtraction or deconvolution, 
when applied to the data, are likely to reveal close binary systems that 
we have missed.

\begin{figure}
\figurenum{14}
\label{binary}
\plotfiddle{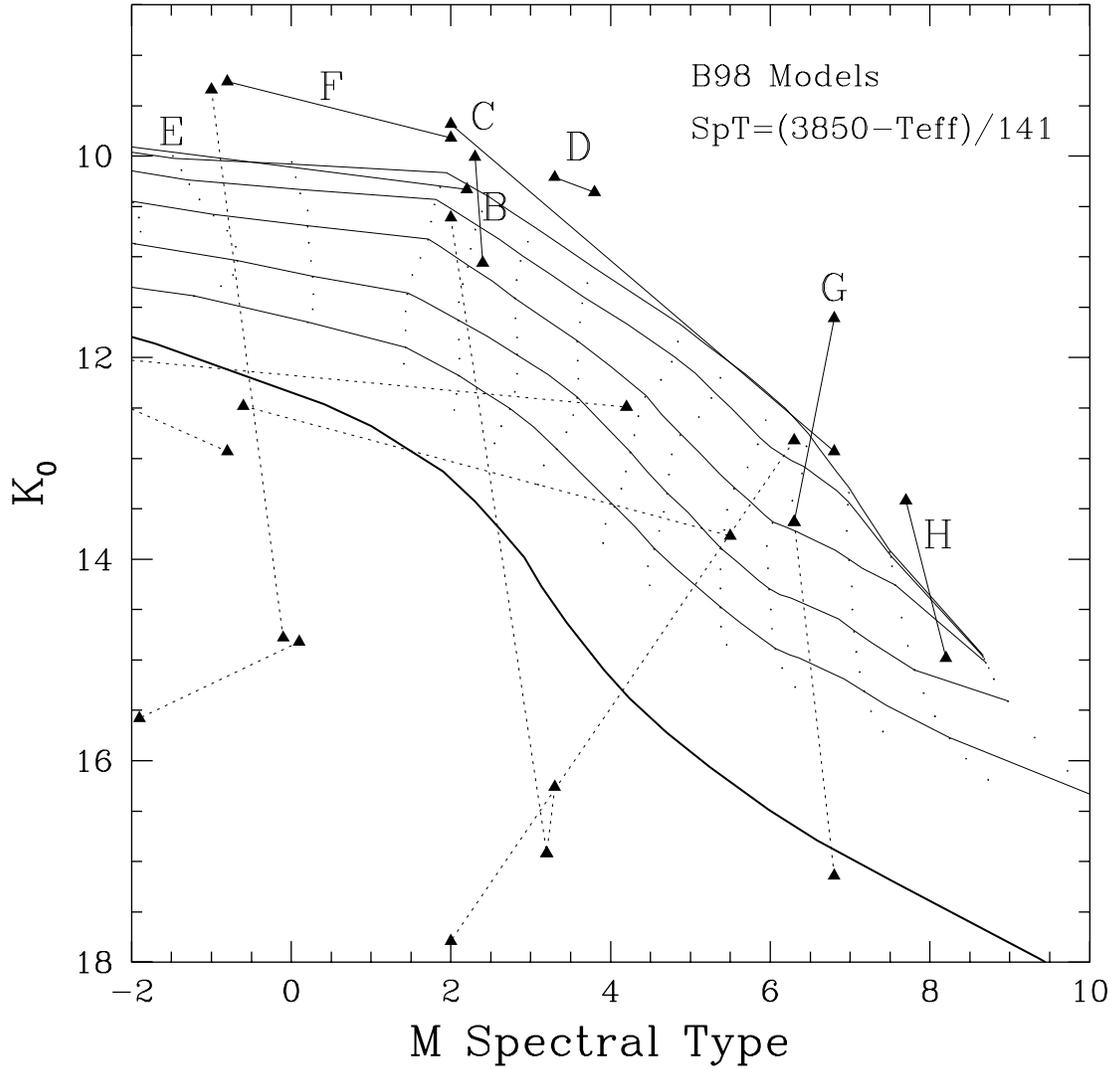}
{6.0in}{0}{75}{75}{-230}{-100}
\caption{Observational HR diagram of pairs of stars separated by 
$< 8\arcsec$.  The pairs are connected by lines.  
Solid (dashed) lines indicate possible (unlikely) physical 
association based on common cluster membership.  
For reference the mass tracks (dotted curves) and 
isochrones (light solid curves)
for the B98 models and \cite{luh99} intermediate temperature scales 
are shown.  The heavy solid line represents the main sequence. 
}
\end{figure}

Table~\ref{binaries} tabulates all of the stars in our sample that 
were found to have a nearest neighbor within $8\arcsec$.  The stars 
have been designated
primary and secondary based on their $K_0$ magnitudes.  The spectral types 
for the G dwarfs are from \cite{luh98} and the other spectral types
are our spectral types as determined in Section 5.  
Figure~\ref{binary} shows the positions of the close pairs 
in the observational HR diagram.  To 
identify the pairs, the components are connected by lines.
Although we were sensitive to separations $\gtrsim 0.8\arcsec$, 
only pairs with separations $> 1.5\arcsec$ were detected.
Based on their locations in the observational HR diagram, 
seven of the close pairs are chance projections of a background star 
close to a cluster member (Fig.~\ref{binary}; dotted lines).
Both components of one pair are background objects.  
Of the 8 candidate cluster binaries 
3 (093-04/093-05; 043-02/043-03; 024-05/024-06)
were previously detected by \cite{duc99} 
in their study of binarity among a sample of 67 IC348 objects.
We also confirm their speculation 
that 083-03 and 023-03 are background 
objects with small projected separations to cluster members.

As shown in Fig.~\ref{binary} (solid lines), several of the 
candidate binary pairs 
have spectral types and $K_0$ magnitudes consistent with a common 
age for the two components.  For the candidate binaries E, F, C, D, 
and H, the lines connecting the two components 
have slopes consistent with the isochrones.
The candidate binary B has a nearly vertical slope. 
However, given our estimate of the uncertainty in the spectral types 
of the binary components, the slope is also highly uncertain, and 
a common age for the binary components cannot be ruled out. 
While the component spectral types for the binary candidate G have similar 
uncertainties, the large separation in magnitude between the two 
components, if each are single stars, makes it unlikely that 
they share a common age.  If, on the other hand, the brighter 
component is an approximate equal mass binary, 
the reduced brightness of each of the two 
stars is more consistent with the evolutionary models, 
and the triple system may be coeval.
If more definitive studies reveal that the binary candidates B and G 
are not coeval, this may indicate that they are not physically related.
Alternatively, a large age difference between the components may 
indicate that the binaries formed through capture.

If we define the binary fraction as the ratio of the number of companions 
detected to the number of targets observed (193 stars), the cluster 
binary fraction in the separation range $0.8\arcsec - 8\arcsec$ 
(240 $-$ 2400 AU) is 8\%.
This is comparable to the result of \cite{duc99} who, 
based on a smaller sample of stars, found a 19\% binary fraction for 
their entire sample;  
half of their binaries fall in the separation range of our study. 
However, there are several important differences between the two studies.
We sample a lower range of primary masses ($\sim 0.015-0.8 M_\odot$) 
than \cite{duc99} ($\sim 0.2 - 2 M_\odot$).  In addition, 
the mass ratios to which we are sensitive are set 
by the magnitude limit of the sample rather than by 
the magnitude difference between the binary components.  
In contrast to \cite{duc99}, who commented on the lack of 
substellar companions, we find candidate substellar companions
(e.g., 022-05) and one candidate substellar binary (H).

\subsection{Low-Mass Cluster Members}

\begin{figure}
\figurenum{15}
\label{lowmass}
\plotfiddle{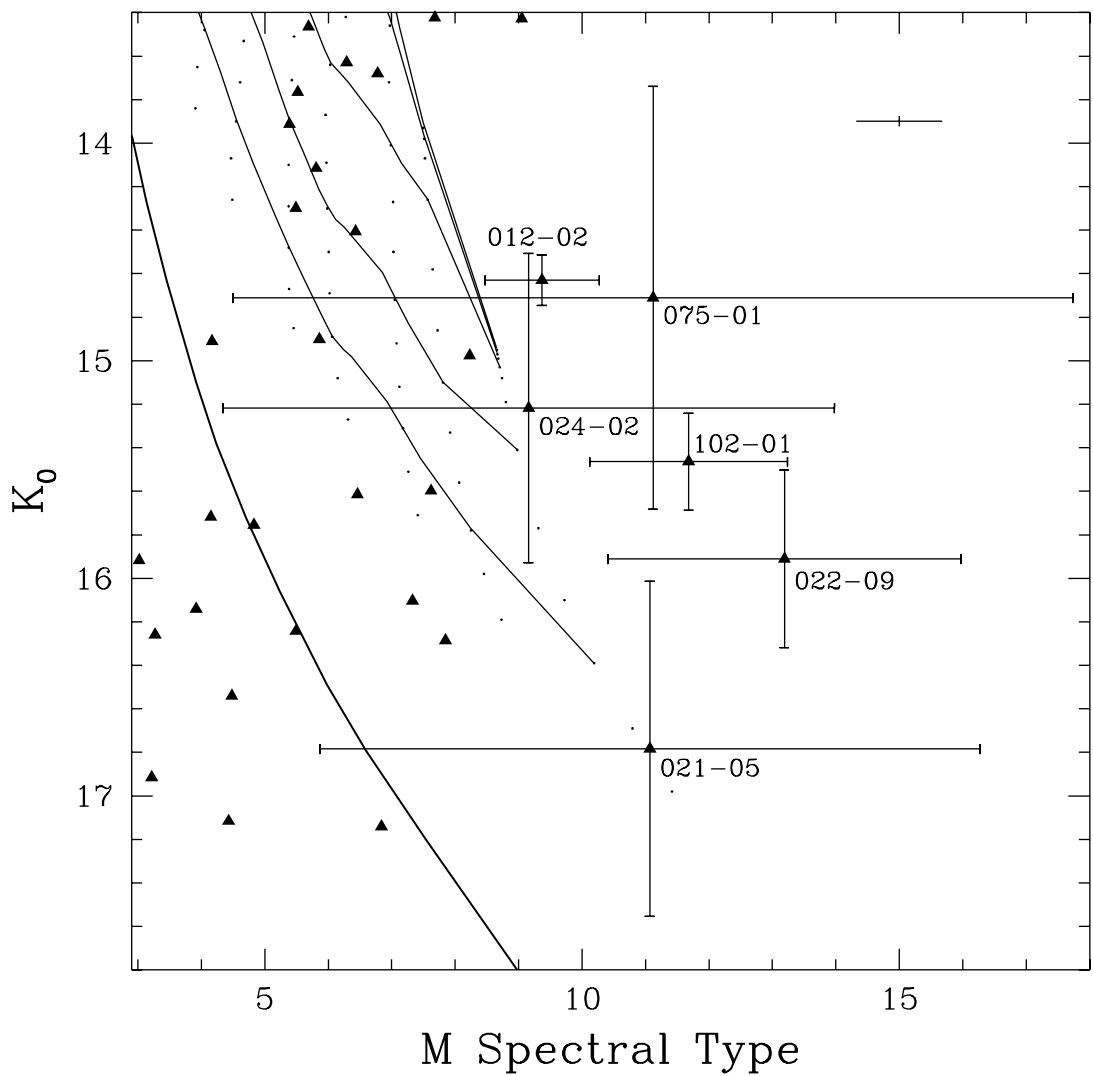}
{6.0in}{0}{75}{75}{-230}{-100}
\caption{ Observational HR diagram highlighting the lowest mass stars 
in the subcluster.  
For reference the mass tracks (dotted curves) and 
isochrones (light solid curves)
for the B98 models and \cite{luh99} intermediate temperature scales 
are shown.  The heavy solid line represents the main sequence. 
The vertical error bars indicate the formal (1--$\sigma$) uncertainty 
in $K_0$ including photometric and extinction uncertainty.  The horizontal 
error bars represent the formal (1--$\sigma$) uncertainty in spectral type.
Typical error bars for the cluster stars are shown in the upper right 
corner.
Note that although the formal uncertainty is assumed to be gaussian, 
it is statistically more likely that 
the stars scatter to earlier rather than later type.}
\end{figure}

The very low-mass cluster population is highlighted in
Figure~\ref{lowmass}.  The 6 objects indicated
have the largest water absorption strengths in the sample,
corresponding to spectral types later than M9, and presumably the lowest masses.
The errors on the derived properties for 3 of the objects
(012-02, 102-01, 022-09) are modest, and imply masses
$\lesssim 0.025 M_\odot$ in the context of both the B98 and DM98 models.
The other 3 objects (024-02, 075-01, and 021-05) are in fact fainter than
our effective limit for accurate spectral typing
($K = 16.5$) and so have spectral type errors $>2.5$ subtypes
(cf. section 7).
Two of these objects, 024-02 ($A_K=1.52$) and 075-01 ($A_K=2.3$),  
are faint due to their large extinctions and are $\sim 5$ and $\sim 7$ 
times more extincted, respectively, than the cluster mean.
Even with the larger errors for these objects, it appears very likely
that all 3 are substellar cluster objects.
However, because of its proximity to the main sequence,
there is a small probability that 021-05 is a background M star.

\subsection{Mass Function}

To estimate a mass function for our sample, we used 
two combinations of evolutionary models and temperature scales: 
the B98 models in combination with the Luhman (1999) intermediate 
temperature scale and the DM98 models in combination with the 
dwarf temperature scale. 
The lower mass limit to which we are complete is determined by 
our spectral typing limit.
As discussed in section 7, we have fairly accurate spectral types 
for all sources to $K=16.5$.  
For a mean cluster reddening of $A_K \simeq 0.3,$ 
this corresponds to $K_0 \simeq 16.2$ or 
$M_K \simeq 8.8$ at the assumed distance of IC348.
Thus, with the DM98 models, we are, for example, complete to 
$0.017 M_\odot$ at the mean extinction of the cluster and 
ages $< 3$ Myr.

For the B98 models, some extrapolation was needed to both 
younger ages ($< 2$ Myr), in order to account for the brighter 
cluster population, and to lower masses ($< 0.025 M_\odot$) 
in order to estimate our mass completeness limit.
In extrapolating below 2 Myr, we used the 1 Myr isochrone from the 
\cite{bar97} models as a guide.  
For the lower masses, 
we used the planetary/brown dwarf evolutionary theory of 
\cite{bur97} to extrapolate the isochrone 
appropriate to the mean age of the subcluster (3 Myr).  
Several similarities between the Burrows et al. and B98 models suggest 
the utility of such an approach.
Like B98, the Burrows et al. theory is non-gray,  
and the evolutionary tracks in the luminosity vs. $T_{\rm eff}$ plane 
at masses $< 0.04 M_\odot$ are qualitatively similar.
Two possible extrapolations are given to illustrate the uncertainty 
in the result. 

In the B98 models, a $0.025 M_\odot$ object at 3 Myr has 
$T_{\rm eff}=2628$K, and $M_K=7.56$.
In comparison, in the Burrows et al. theory, a $0.025 M_\odot$ object 
at 3 Myr is slightly hotter ($T_{\rm eff}$=2735K) but has a comparable 
absolute $K$ magnitude ($M_K=7.6$ assuming BC$_K = 3.0$);  
a 3 Myr old object that is 1.2 magnitudes 
fainter ($M_K=8.8$) has an effective temperature $\simeq 300$K cooler 
and is $0.011 M_\odot$ lower in mass.  
Applying the same mass and temperature differentials to 
the 3 Myr old, $0.025 M_\odot$ object from B98 
implies that a 3 Myr old, $M_K=8.8$ object 
in the B98 theory has $T_{\rm eff}=2330$K and a mass of $0.014 M_\odot$.

As an alternate estimate, we can extrapolate the 3 Myr isochrone 
based on a match in $T_{\rm eff}$ rather than mass.  As described above, 
the effective temperature of a $0.025 M_\odot$, 3 Myr old object 
in B98 theory is 2628K. 
From the $T_{\rm eff}=2628$K point in the 3 Myr isochrone of the 
Burrows et al. models, $\Delta M_K=1.2$ corresponds 
to a change in temperature and mass of $\Delta T_{\rm eff}=-425$K 
and $\Delta m=-0.010 M_\odot$.
Applying these mass and temperature differentials to 
the 3 Myr old, $0.025 M_\odot$ object from B98 
implies an effective temperature of $2200$K and mass $0.015 M_\odot$
for a 3 Myr old, $M_K=8.8$ object.  
Thus, with either estimate, our spectral typing limit of 
$M_K=8.8$ corresponds to a mass completeness limit of $\sim 0.015 M_\odot$ 
at the average age and reddening of the cluster members.
The effective temperature appropriate to this mass limit is less 
certain. 

Formally, the appropriate effective temperature affects our estimate 
of the lower limit to the final mass bin of our sample. 
Note, however, that our spectral typing limit 
of $0.015 M_\odot$ is close to the deuterium burning limit (Burrows et al. 1993, 
Saumon et al. 1996) and in the age range in which objects 
fade fairly rapidly with age.  
For example, in the Burrows et al. models, a 3 Myr old, $0.010 M_\odot$ 
object is half as luminous as a $0.015 M_\odot$ object at the same age. 
Given the rapid fading, it is unlikely that we have detected objects 
much less massive than $0.015 M_\odot$, 
which we adopt as the lower limit 
of the final mass bin of the sample.  
Note that our spectral typing limit of $K=16.5$ implies that 
we have somewhat underestimated the population of the final mass bin if 
that bin is characterized by the same spread in age and reddening that 
is measured at higher masses.

\begin{figure}
\figurenum{16}
\label{massfunc}
\plotfiddle{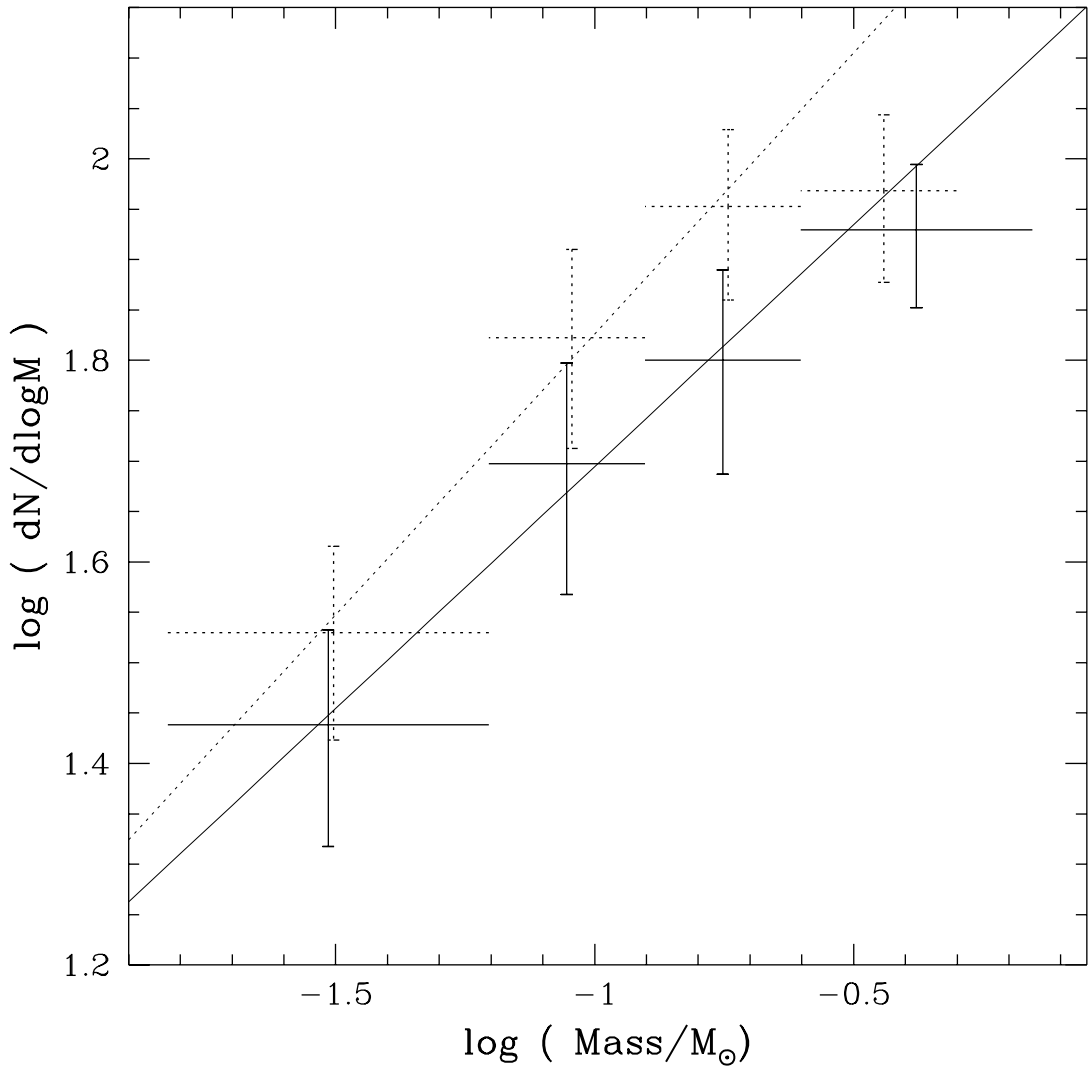}
{6.0in}{0}{75}{75}{-230}{-100}
\caption{The mass function of the IC348a subcluster as derived from 
two combinations of evolutionary tracks and temperature scales: 
the B98 models in combination with the \cite{luh99} intermediate 
temperature scale (solid) and the DM98 models in combination 
with the dwarf temperature scale (dotted).  The horizontal 
lines indicate the width of the mass bins.  The vertical error bars 
include only the $\sqrt N$ errors associated with the counting 
statistics.  Fits to the 3 lowest mass bins indicate 
$dN/d\log M \propto M^{0.5}$ (B98) and 
$dN/d\log M \propto M^{0.6}$ (DM98).
}
\end{figure}

The mass functions for the age range $0-10$ Myr 
that result from the assumptions and extrapolations 
discussed above are shown in Fig.~\ref{massfunc}.
The result for both the B98 models (solid symbols) and DM98 models 
(dotted symbols) are shown.
Note that the objects indicated previously as potential background 
objects ($K_0=15.5-16.5$, M6$-$M8) are not included in the
mass function for the B98 models, whereas some are included in the 
mass function for the DM models.  Since these objects represent only 
a small fraction of the objects in each bin, whether or not these are 
included as members makes little difference to the slope of the 
mass function.

The DM98 models indicate a flattening at $\sim 0.25 M_\odot,$ whereas 
the B98 models imply an approximately constant slope over the entire 
mass range $0.7-0.015 M_\odot.$ In either case, 
the mass function appears to decrease from $\sim 0.25 M_\odot$, 
through the hydrogen burning limit ($\sim 0.08 M_\odot$), 
down to the deuterium burning limit ($\sim 0.015 M_\odot$).
The slope of the mass function in this range is consistent with 
$dN/d\log M \propto M^{0.5}$ for B98 and 
$dN/d\log M \propto M^{0.6}$ for DM98. 
The slow, approximately continuous decrease in the mass function in this 
interval differs from the result obtained by Hillenbrand (1997) for the 
Orion Nebula Cluster.  The sharp fall off in the Orion Nebula Cluster 
mass function below 
$\sim 0.2 M_\odot$ ($dN/d\log M \propto M^{2.5}$) is not reproduced here.
Instead, we find that the slope of the IC348 mass function is more 
similar to that derived for the Pleiades in the mass range $0.3-0.04 M_\odot,$ 
$dN/d\log M \propto M^{0.4}$ (\cite{bou98}).
The slope is similar to that inferred for the 
substellar population of the solar neighborhood from 2MASS and DENIS data.
As determined by Reid et al.\ (1999), the observed properties of the 
local L dwarf population are consistent with a mass function 
$dN/d\log M \propto M^{\alpha},$ with $\alpha\sim$ -1 to 0, although a 
mass function similar to that for IC348 is not strongly precluded 
especially given the uncertainty in the age distribution of objects 
in the solar neighborhood.

Given the low masses to which we are sensitive, it is also interesting 
to compare our result to the mass function that is emerging for 
companions to nearby solar-type (G$-$KV) stars at separations $<5$ AU
(e.g., \cite{mar00}).
While initial results indicated that the substellar companion mass 
function might be a smooth continuation of the stellar companion mass 
function (e.g., $dN/d\log M \propto M^{0.6}$; \cite{may98}), 
proper motion data from Hipparcos have revealed that 
a significant fraction of companions in the $0.015-0.08 M_\odot$ range 
are low inclination systems, and hence have larger (stellar or near-stellar) 
masses (\cite{mar00}; Halbwachs et al.\ 2000). 
When corrected for these low inclination systems, the 
companion mass function appears to be characterized by a marked deficit 
in the $0.015-0.08 M_\odot$ mass range (the ``brown dwarf desert''; 
\cite{mar00}; Halbwachs et al.\ 2000).
In contrast, the mass function for IC348 appears to decrease 
continuously through the stellar/substellar boundary and 
the mass range $0.08 - 0.015 M_\odot$.

The low mass end of the IC348 sample extends into the mass range ($10-20 M_J$)
in which objects transition from higher mass objects that burn deuterium 
early in their evolution to lower mass objects that are incapable of 
deuterium burning due to the onset of electron degeneracy pressure 
during the contraction phase 
(e.g., \cite{gro74}; \cite{bur93}).  
According to the calculations of Saumon et al.\ (1996), 
$\sim 15 M_J$ objects deplete their deuterium abundances by a factor 
of 2 after 30 Myr of evolution, while objects $\le 12 M_J$ retain all of 
their initial deuterium and derive no luminosity from thermonuclear 
fusion at any point in their evolution. 
They suggest the deuterium burning limit as a possible interpretive 
boundary between objects that are regarded as brown dwarfs and 
those regarded as planets. 

If we assume that the hydrogen and deuterium burning mass limits 
delimit the brown dwarf population,  
with either the DM98 or B98 models, 
we have fully sampled the brown dwarf population, at ages up to the 
mean age of the subcluster and extinctions up to the cluster average.
Thus, we can conclude with near certainty that the fraction of 
the subcluster mass contributed by brown dwarfs is low, only a few 
percent of the cluster mass.  
With the B98 tracks, we find a total of $\sim 22$ cluster substellar
candidates which represents a significant fraction, $\sim 20$\%, of all
cluster M dwarfs by number, but only a small fraction, $\sim 4$\%, by
mass.  For comparison, with the DM98 tracks, we find $\sim 30$ cluster
substellar candidates which represents $\sim 30$\% of all cluster 
M dwarfs by number and $\sim 6$\% by mass.

These limits on the substellar contribution to the total cluster mass 
have interesting implications when compared with current limits  
placed by microlensing studies on the substellar content of the 
Galactic halo.  Based on the search for microlensing toward the LMC  
the current EROS limits on the fraction of the halo mass that resides 
in brown dwarf mass objects is $\lesssim$ 10\% (Lasserre et al.\ 2000). 
Scaling our results for IC348 by the stellar fraction of the halo 
mass ($\sim 1$ \%), we find that if the halo has the same IMF as 
IC348, then substellar objects contribute negligibly to the halo mass 
($< 0.1$\%).  The several orders of magnitude difference between these 
limits leaves room for some interesting possibilities.  If future 
microlensing results find confirmation for a halo mass fraction of even 
$\sim 1$ \% in substellar objects, that would indicate that low mass 
star formation in the halo proceeded significantly differently from 
that currently occuring in Galactic clusters.

What do the IC348 results tell us about the star formation process?
The absence of structure in the mass function at the hydrogen burning 
limit (e.g., a turnover) is perhaps expected.  It is difficult to 
imagine how hydrogen burning, which demarcates the end of the 
pre-main sequence phase, could influence the determination of 
stellar masses, an outcome which is probably determined at much 
earlier times.

We also find no obvious feature in the IMF at the deuterium burning 
limit (e.g., a strong increase or decrease), 
a potentially more relevant mass scale for star formation since 
deuterium burning occurs at pre-main-sequence ages.
This result may appear puzzling in the context of some current 
theories for the origin of stellar masses.  
For example, in a canonical theory of the formation of solar-type 
stars, it is the onset of deuterium burning that is believed to 
set in motion the sequence of events by which a star comes to have 
a role in determining its own mass.  The onset of deuterium burning 
first induces a fully convective stellar interior.  The convective 
interior, combined with the rapid stellar rotation that is likely to 
result from the  accretion of angular momentum along with mass, is 
believed to generate a strong stellar magnetic field.  The strong field 
is, in turn, believed to drive a magnetocentrifugal wind that ultimately 
sweeps away the cloud from which the star formed and possibly reverses 
the infall itself, thereby helping to limit the mass of the star.  The 
self-deterministic aspect of such a mass-limiting wind is a 
critical element in explanations for the generic origin of 
stellar masses (e.g., \cite{shu95}) and some theories of the IMF 
(e.g., \cite{ada96}).

In this picture, as masses close to the deuterium burning limit are 
approached, one might expect that, the deuterium burning trigger being 
absent, low mass objects might not be able to reverse the infall and, 
consequently, it would be difficult to produce any objects of such 
low mass.  This appears to be inaccurate both theoretically and 
observationally.  Not only are young objects in this mass range fully 
convective without the aid of deuterium burning (Burrows, personal 
communication) and may thereby generate magnetic fields in advance 
of or in the absence of deuterium burning, but we also find no deficit 
of objects near the deuterium burning limit.  This nevertheless raises 
the important question of what physical processes determine the masses 
of objects much below a solar mass.  

Fragmentation is one possibly significant process at this mass scale. 
Coindentally, our survey mass limit is close to the characteristic mass 
for opacity-limited fragmentation under the low temperature, chemically 
enriched conditions current prevailing in molecular clouds 
($\sim 0.01 M_\odot$; e.g., Silk 1977).  In this picture, if cooling is 
efficient as collapse proceeds, the inverse dependence of the Jeans 
mass on density leads to fragmentation on increasingly small scales as 
collapse continues, halting only when objects become optically thick 
to their own radiation and the cooling efficiency is thereby impaired.  
If the characteristically low mass objects that form as the result of this 
process represent the ``seeds'' of star formation from which more massive 
objects must grow, we might expect to find a large number of objects 
with this mass.  Perhaps significantly, we find no such large excess, 
but rather a smooth continuation from the stellar mass regime down to 
this mass scale.  This implies that if fragmentation plays an important 
role in the formation of stars and brown dwarfs, that the subsequent 
events (e.g., merging, accretion) are efficient at erasing the 
characteristic mass scale for fragmentation.  
Future IMF studies that probe masses below the characteristic fragmentation 
mass can provide more stringent constraints on the role of fragmentation 
in the star formation process.

\section{Summary and Conclusions}

Using HST NICMOS narrow band imaging, we have measured the 
1.9 $\mu$m water band strengths of low-mass objects in the IC348a 
subcluster.  With the magnitudes and spectral types 
thereby obtained, we are able to separate 
cluster members from background objects.
Comparisons with recent evolutionary tracks (B98, DM98) imply 
that our study probes a mass range extending from low-mass stars 
($\lesssim 0.7 M_\odot$)
down to the bottom of the deuterium burning main sequence 
($\gtrsim 0.015 M_\odot$).
The mean age of the subcluster is 3 Myr with the B98 tracks 
and 1 Myr with the DM98 tracks.
These results are subject to uncertainties in the evolutionary tracks 
and the appropriate conversions between theoretical 
($L_*$, $T_{\rm eff}$) and observed 
(e.g., spectral types, magnitudes) 
quantities which remain somewhat uncertain.
We also confirm an age spread to the cluster, as found previously 
(\cite{lad95}; \cite{her98}; \cite{luh98}),  
from $<1$ to $10-20$ Myr.

Assuming that the hydrogen- and deuterium-burning mass limits 
delimit the brown dwarf population, we have fully sampled the brown 
dwarf population at ages up to 3 Myr 
and extinctions up to the cluster average ($A_K=0.3$).  
We find $\sim 20-30$ cluster substellar candidates (depending on 
the choice of evolutionary tracks) which represents a significant 
fraction, $\sim 25$\%, of all cluster M dwarfs by number, 
but only a small fraction, $\sim 5$\%, by mass.  
The mass function derived for the subcluster, 
$dN/d\log M \propto M^{0.5},$ is similar to that recently obtained 
for the Pleiades over a more limited mass range (\cite{bou98}), 
and apparently less abundant in low mass objects than the 
local field population (\cite{rei99}). 
In contrast, the derived mass function appears significantly more abundant 
in brown dwarfs than the mass function of companions 
to nearby solar-type stars at separations $<5$ AU (\cite{mar00}). 

The apparent difference may indicate 
that substellar objects form more readily in isolation than as 
companions.  Alternatively, the difference may represent the result 
of evolutionary effects such as 
accretion (by the star) or dynamical ejection, which will 
tend to deplete the companion population and, in the latter case, 
contribute low mass objects to the local field population.  
Given the population statistics from precision radial velocity studies,  
if these evolutionary mechanisms are the underlying physical cause for
the different IMFs, they must preferentially deplete the brown dwarf 
population compared to the lower mass planetary companion population,
which appears to be present in significant numbers.

More generally, we find that the imaging photometric 
technique used in this study is a potentially powerful approach to 
the study of low mass populations in young clusters.  
As demonstrated here, it is possible to study a large range in mass 
($\sim 0.5 - 0.015 M_\odot,$ a factor of $> 30$ in mass) with a 
single technique.
To summarize, the utility of this approach derives from 
the multi-object approach inherent in a filter photometric method; 
the sensitivity of the index due to the
rapid variation of the water band strength  with late-M spectral type; 
the approximate orthogonality of the reddening vector 
to the variation with spectral type so that 
reddening errors do not introduce significant spectral type errors; 
and the long wavelength of the index 
which improves the sampling of embedded populations.   

To stress this latter point, we can consider the depth to which 
one would have to carry out spectroscopy in the $I$-band
to recover similar information for IC348.
Our completeness limit for spectral typing is $K \simeq 16.5$.
With this level of completeness, we have sampled a significant
fraction of the low-mass cluster population.  For example, to
$A_K=0.5$, the B98 and \cite{bur97} tracks imply that at ages 
of 20, 3, and 1 Myr, we are
complete to 35, 16, and 9 $M_J$.  For a more extreme extinction of
$A_K=2$, the B98 model suggests that at ages of 20, 3, and 1 Myr, we 
are complete to 100, 32, and 25 $M_J$.  In contrast, for a spectroscopic 
study in the $I$-band, $A_I/A_K\simeq 5$ and for the late spectral 
types probed in the present study, $I-K \sim 4.5$.
Consequently, for extinctions of $A_K=$0.5 and 2,
the corresponding limiting magnitude is $I=23$ and 29.
In contrast, optical spectral typing with existing 10-m telescopes
is currently limited to sources brighter than  $I\simeq 19.5.$

Although the present study made use of narrow band filters and 
the ability to work above the Earth's atmosphere with HST,  
the technique used here might find useful extrapolation to both broader 
filters and to ground-based observations.  
With broader filters, it would be possible to study objects 
at lower, planetary masses, as well as more distant, richer  
clusters where the spatial multiplexing advantage of a filter photometric 
technique could be used to better advantage. 
We will explore these possibilities in a future paper (\cite{tie00}).

\acknowledgements

We are grateful to 
Nick Bernstein and Alex Storrs for their extensive help in getting 
our program scheduled and executed; 
to Matt Lallo and Russ Makidon for their explanation of HST pointing errors; 
to Marcia Rieke and Paul Martini for their help with the photometric 
calibration; 
to Wolfram Freudling for advice on NICMOSlook; 
to Tod Lauer who helped us investigate the impact of intra-pixel sensitivity 
on our photometry; 
to France Allard and Peter Hauschildt for sharing their atmosphere models; 
to Adam Burrows who shared his brown dwarf evolutionary models; 
to Martin Cohen who helped us estimate the background population; 
and to Kevin Luhman for useful advice and sharing his IC348 results with 
us in advance of publication. 
We are also grateful to Charles Lada and Tom Greene for useful discussions 
regarding this project, and to Arjun Dey and the anonymous referee 
whose comments significantly improved the manuscript.
Support for this work was provided by NASA through grant number 
GO-07322.02-96A
from the Space Telescope Science Institute, which is operated by AURA, Inc., 
under NASA contract NAS5-26555.

\clearpage

\clearpage

\clearpage

\begin{deluxetable}{llrrrrrrrrl}
\tablecaption{Standard Stars\label{standards}}
\tablewidth{0pt}
\footnotesize
\tablenum{1}
\tablehead{
     \colhead{ID}
   & {Spectral}
   & {$K$\tablenotemark{b}}
   & {$A_V$}
   & \colhead{F166}
   & \colhead{err(F166)}
   & \colhead{F190}
   & \colhead{err(F190)}
   & \colhead{F215}
   & \colhead{err(F215)}
   & \colhead{Notes} \\[.3ex]
\colhead{}
   & \colhead{Type\tablenotemark{a}}
   & \colhead{}
   & \colhead{}
   & \colhead{(Jy)}
   & \colhead{(Jy)}
   & \colhead{(Jy)}
   & \colhead{(Jy)}
   & \colhead{(Jy)}
   & \colhead{(Jy)}
   & \colhead{}
   }
\tablenotetext{a}{Unless otherwise noted, spectral types are from Kirkpatrick et al. (1991)}
\tablenotetext{b}{Unless otherwise noted, $K$ photometry are from Leggett (1992)}
\tablenotetext{c}{Spectral type from Keenan \& McNeil (1989)}
\tablenotetext{d}{Spectral type from Henry et al. (1994)}
\tablenotetext{e}{Spectral type from Walter et al. (1988)}
\tablenotetext{f}{WTTS; $K$ photometry and extinction from Kenyon \& Hartmann (1995)}
\tablenotetext{g}{Spectral type from Gliese (1969)}
\tablenotetext{h}{Counts were above the saturation limit in the last read.  Photometric values are based on counts corrected for saturation by the {\it calnica} pipeline.}
\tablenotetext{i}{Spectral type from Hartmann et al. (1991)}
\tablenotetext{j}{Spectral type from Kirkpatrick et al. (1995)}
\tablenotetext{k}{$K$ photometry from Forrest et al. (1988) and Henry \& Kirkpatrick (1990)}
\startdata
Gl764.1A	&	K2V	&	\nodata	&	\nodata	&	2.60E+0	&	5.90E-3	&	2.17E+0	&	4.66E-3	&	1.85E+0	&	4.38E-3	&	c	\nl
Gl795	&	K5V	&	\nodata	&	\nodata	&	1.16E+1	&	2.21E-2	&	9.23E+0	&	1.45E-2	&	8.34E+0	&	1.52E-2	&		\nl
Gl764.1B	&	K7V	&	\nodata	&	\nodata	&	1.51E+0	&	6.60E-3	&	1.24E+0	&	5.84E-3	&	1.11E+0	&	5.58E-3	&		\nl
Gl328	&	M0V	&	6.42	&	\nodata	&	2.57E+0	&	9.54E-3	&	2.02E+0	&	7.36E-3	&	1.83E+0	&	7.89E-3	&		\nl
Gl908	&	M1V	&	5.05	&	\nodata	&	8.14E+0	&	1.92E-2	&	6.10E+0	&	1.54E-2	&	5.95E+0	&	1.52E-2	&	d	\nl
HBC362	&	M2V	&	10.06	&	0.28	&	9.35E-2	&	1.79E-3	&	6.58E-2	&	1.43E-3	&	6.24E-2	&	1.41E-3	&	e,f	\nl
Gl195A	&	M2V	&	6.01	&	\nodata	&	3.18E+0	&	9.82E-3	&	2.48E+0	&	8.15E-3	&	2.32E+0	&	8.14E-3	&	g	\nl
Gl569A	&	M3V	&	\nodata	&	\nodata	&	3.70E+0	&	6.16E-3	&	3.10E+0	&	4.18E-3	&	3.06E+0	&	4.36E-3	&	h	\nl
HBC360	&	M3V	&	9.98	&	0.28	&	9.69E-2	&	1.91E-3	&	7.25E-2	&	1.43E-3	&	6.82E-2	&	1.47E-3	&	e,f	\nl
HBC361	&	M3V	&	10.11	&	0.28	&	8.12E-2	&	2.24E-3	&	5.90E-2	&	1.68E-3	&	6.22E-2	&	1.72E-3	&	e,f	\nl
Gl388	&	M3V	&	4.61	&	\nodata	&	1.07E+1	&	2.13E-2	&	8.61E+0	&	1.45E-2	&	8.93E+0	&	1.56E-2	&	d	\nl
Gl896A	&	M3.5V	&	5.58	&	\nodata	&	6.11E+0	&	1.73E-2	&	4.57E+0	&	1.36E-2	&	4.57E+0	&	1.45E-2	&	d	\nl
Gl213	&	M4V	&	6.37	&	\nodata	&	2.24E+0	&	9.29E-3	&	1.68E+0	&	7.03E-3	&	1.78E+0	&	7.34E-3	&		\nl
Gl83.1	&	M4.5V	&	6.67	&	\nodata	&	1.68E+0	&	5.63E-3	&	1.28E+0	&	4.69E-3	&	1.38E+0	&	5.01E-3	&		\nl
Gl896B	&	M4.5V	&	\nodata	&	\nodata	&	1.56E+0	&	1.90E-2	&	1.15E+0	&	1.79E-2	&	1.21E+0	&	1.68E-2	&	d	\nl
J1-4423	&	M5V	&	10.43	&	0.97	&	5.87E-2	&	1.37E-3	&	4.04E-2	&	1.09E-3	&	4.48E-2	&	1.14E-3	&	i,f	\nl
Gl406	&	M6V	&	6.08	&	\nodata	&	2.94E+0	&	9.76E-3	&	2.13E+0	&	7.45E-3	&	2.39E+0	&	7.92E-3	&		\nl
GJ1111	&	M6.5V	&	7.26	&	\nodata	&	9.37E-1	&	6.08E-3	&	6.68E-1	&	4.50E-3	&	7.94E-1	&	4.75E-3	&	d	\nl
LHS3003	&	M7V	&	8.93	&	\nodata	&	1.74E-1	&	5.01E-3	&	1.41E-1	&	4.13E-3	&	1.72E-1	&	4.34E-3	&	j	\nl
VB8	&	M7V	&	8.81	&	\nodata	&	2.25E-1	&	4.48E-3	&	1.51E-1	&	3.47E-3	&	2.05E-1	&	3.78E-3	&		\nl
VB10	&	M8V	&	8.80	&	\nodata	&	2.00E-1	&	5.78E-3	&	1.50E-1	&	4.64E-3	&	1.97E-1	&	4.75E-3	&		\nl
Gl569B	&	M8.5V	&	9.56	&	\nodata	&	9.53E-2	&	4.30E-3	&	7.66E-2	&	3.38E-3	&	9.49E-2	&	3.43E-3	&	k	\nl
LHS2924	&	M9V	&	10.69	&	\nodata	&	3.53E-2	&	7.22E-4	&	2.53E-2	&	5.77E-4	&	3.47E-2	&	5.95E-4	&		\nl
\enddata
\end{deluxetable}

\clearpage

\begin{deluxetable}{llc}
\tablecaption{Log of Observations\label{log}}
\tablewidth{0pt}
\tablenum{2}
\tablehead{
     \colhead{}
   & \colhead{Date}
   & \colhead{Exposure} \\[.3ex]
\colhead{Object\tablenotemark{a}}
   & \colhead{(yymmdd)}
   & \colhead{(num. $\times$ sec.)}
   } 
\tablenotetext{a}{Each object was observed in all three narrow bands filters, F166N, F190N, and F215N.  Standard
stars were also observed in the G141 and G206 grisms.}
\startdata
\multicolumn{3}{c}{Standards} \nl
\hline
Gl896AB    & 980112 & $ 2 \times 0.60 $ \nl
Gl569AB    & 980113 & $ 2 \times 1.99 $ \nl
LHS3003    & 980113 & $ 2 \times 1.21 $ \nl
GJ1111     & 980117 & $ 2 \times 1.21 $ \nl
Gl213      & 980117 & $ 2 \times 0.91 $ \nl
Gl195A     & 980118 & $ 2 \times 0.91 $ \nl
Gl328      & 980118 & $ 2 \times 0.91 $ \nl
Gl83.1     & 980119 & $ 2 \times 1.21 $ \nl
Gl388      & 980123 & $ 2 \times 0.30 $ \nl
Gl406      & 980124 & $ 2 \times 0.91 $ \nl
HBC362     & 980125 & $ 2 \times 2.99 $ \nl
VB8        & 980125 & $ 3 \times 1.21 $ \nl
HBC360/361 & 980127 & $ 2 \times 2.99 $ \nl
LHS2924    & 980127 & $ 1 \times 0.20 $ \nl
J14423     & 980130 & $ 2 \times 3.98 $ \nl
Gl795      & 980616 & $ 2 \times 0.60 $ \nl
Gl764.1AB  & 980616 & $ 2 \times 1.21 $ \nl
~~~~"      & ~~~"   & $ 1 \times 0.20 $ \nl
Gl908      & 980619 & $ 2 \times 0.30 $ \nl
Gl699      & 980619 & $ 2 \times 0.30 $ \nl
VB10       & 980621 & $ 3 \times 1.21 $ \nl
\hline 
\vspace{-1.3ex} \nl
\multicolumn{3}{c}{IC348 Fields} \nl
\hline
011 -- 035  & 980114 & $ 2 \times 128.00 $ \nl
041 -- 065  & 980115 & $ 2 \times 128.00 $ \nl
071 -- 095  & 980116 & $ 2 \times 128.00 $ \nl
101 -- 105  & 981022 & $ 2 \times 128.00 $ \nl
\enddata
\end{deluxetable}

\clearpage

\begin{deluxetable}{lrrr}
\tablecaption{Photometric Calibration Constants\label{photcal}}
\tablewidth{0pt}
\tablenum{3}
\tablehead{
     \colhead{}
   & \colhead{$\lambda_{eff}$}
   & \colhead{Jy}
   & \colhead{0 mag\tablenotemark{a}} \\[.2ex]
\cline{3-3}\\[-2.5ex]
\colhead{Filter}
   & \colhead{($\mu$m)}
   & \colhead{ADU/s}
   & \colhead{(Jy)}
   }
\tablenotetext{a}{Flux based on Vega zero point}
\startdata
F166N & 1.658 & 5.911E-05 & 1010  \nl
F190N & 1.900 & 4.920E-05 &  808  \nl
F215N & 2.149 & 4.896E-05 &  689  \nl
\enddata
\end{deluxetable}

\clearpage

\begin{deluxetable}{lccccc}
\tablecaption{Observed/Model Background\label{cohenbck}}
\tablewidth{0pt}
\tablenum{5}
\tablehead{
	\colhead{SpT}
	& \multicolumn{5}{c}{$K_0$ bins} \\[0.5ex] \cline{2-6}\\[-2.0ex]
	& \colhead{0-13}
	& \colhead{13-14}
	& \colhead{14-15}
	& \colhead{15-16}
	& \colhead{16-17}
}
\startdata
$<$M3 	   &  7/2.1  & 4/6.4  & 13/12.4 & 21/19.3 & 7/24.0 \nl
M4--5 	   &  0/0.0  & 0/0.2  &  0/0.5  &  2/1.4  & 4/4.1  \nl
$>$M5 	   &  0/0.0  & 0/0.0  &  0/0.1  &  0/0.4  & 1/1.0  \\[0.5ex] 
\hline\\[-2.0ex]
Total      &  7/2.1  & 4/6.6  & 13/13.0 & 23/21.1 & 12/29.1 \nl
\enddata
\end{deluxetable}

\clearpage

\begin{deluxetable}{rrrrrllrrl}
\tablecaption{Objects with Separations $<8\arcsec$\label{binaries}}
\tablewidth{0pt}
\tablenum{6}
\tablehead{
     \colhead{Primary}
   & \colhead{Secondary}
   & \colhead{Sep.}
   & \colhead{$K_0$}
   & \colhead{$K_0$}
   & \colhead{Sp.T.}
   & \colhead{Sp.T.}
   & \colhead{$A_K$}
   & \colhead{$A_K$}
   & \colhead{ID in} \\[.3ex]
\colhead{(p)}
   & \colhead{(s)}
   & \colhead{$(\arcsec)$}
   & \colhead{(p)}
   & \colhead{(s)}
   & \colhead{(p)}
   & \colhead{(s)}
   & \colhead{(p)}
   & \colhead{(s)}
   & \colhead{Fig.~14}
   }
\startdata
$084$-01 & 084-02 & 1.51 & 9.13 & 9.08 & G0.0 & G0.0 & 0.56 & 0.42 &   \nl
$093$-04 & 093-05 & 1.53 & 10.01 & 11.06 & M2.3 & M2.4 & 0.98 & 0.26 & B  \nl
$022$-04 & 022-05 & 1.98 & 9.68 & 12.93 & M2.0 & M6.8 & 1.12 & 0.53 & C  \nl
$062$-04 & 062-05 & 2.86 & 10.21 & 10.36 & M3.3 & M3.8 & 0.54 & 0.43 & D  \nl
$043$-03 & 043-02 & 5.02 & 9.71 & 10.33 & K6.0 & M2.2 & 0.10 & 0.30 & E  \nl
$024$-06 & 024-05 & 5.56 & 9.26 & 9.82 & K9.2 & M2.0 & 0.19 & 0.00 & F  \nl
$014$-04 & 014-05 & 6.67 & 11.61 & 13.63 & M6.8 & M6.3 & 0.22 & 0.04 & G  \nl
$013$-06 & 013-04 & 7.80 & 13.42 & 14.98 & M7.7 & M8.2 & 0.26 & 0.00 & H  \nl
 &  &  &  &  &  &  &  &  &   \nl
$083$-03 & 083-02 & 2.92 & 9.34 & 14.78 & K9.0 & K9.9 & 0.82 & 1.23 &   \nl
$014$-05 & 014-06 & 3.30 & 13.63 & 17.14 & M6.3 & M6.8 & 0.04 & 0.62 &   \nl
$055$-02 & 055-03 & 5.16 & 12.82 & 17.79 & M6.3 & M2.0 & 0.24 & 0.00 &   \nl
$052$-02 & 052-03 & 6.21 & 16.26 & 16.92 & M3.3 & M3.2 & 0.38 & 0.00 &   \nl
$021$-07 & 022-06 & 6.35 & 11.28 & 12.49 & G8.0 & M4.2 & 0.81 & 0.00 &   \nl
$052$-04 & 052-03 & 6.64 & 10.61 & 16.92 & M2.0 & M3.2 & 0.71 & 0.00 &   \nl
$022$-01 & 022-02 & 6.81 & 14.82 & 15.58 & M0.1 & K8.1 & 0.83 & 1.23 &   \nl
$082$-04 & 094-04 & 6.87 & 12.48 & 13.77 & K9.4 & M5.5 & 1.04 & 0.84 &   \nl
$023$-03 & 023-02 & 7.44 & 9.69 & 12.93 & K0.0 & K9.2 & 0.04 & 1.16 &   \nl
\enddata
\end{deluxetable}

\begin{table}
\tablenum{4}
\label{astrometry}
\end{table}


\begin{thebibliography}{}

\bibitem[Adams \& Fatuzzo 1996]{ada96} Adams, F.C., \& Fatuzzo, M. 1996, 
ApJ, 464, 256

\bibitem[Ali et al.\ 1995]{ali95} Ali, B., Carr, J. S., Depoy, D. L.,
Frogel, J. A., \& Sellgren, K. 1995, AJ, 110, 2415

\bibitem[Allard et al.\ 1997]{all97} Allard, F., Hauschildt, P.H., 
Alexander, D.R., \& Starrfield, S. 1997, ARA\&A 35, 137

\bibitem[Allard 1998a]{all98a}
Allard, F. 1998a, in ``Brown Dwarfs and Extra-solar Planets'', 
ASP Conf. Ser. 134, eds. R. Rebolo, E. Mart\'{\i}n, \& M. R. Zapatero Osorio, p. 370

\bibitem[Allard 1998b]{all98b} Allard, F. 1998b, 
``Modelling M-dwarfs Atmospheres'', 
in Proceedings of the Euroconference on ``Very Low-Mass Stars and Brown 
Dwarfs in Stellar Clusters and Associations'', Los Cancajos, La Palma, 
Spain, May 11-15, 1998.

\bibitem[Allard et al.\ 2000]{all00} Allard et al.\ 2000, in preparation

\bibitem[Baraffe et al.\ (1997)]{bar97}
Baraffe, I., Chabrier, G., Allard, F., \& Hauschildt P.H. 1997, A\&A, 327, 1054

\bibitem[Baraffe et al.\ 1998]{bar98}
Baraffe, I., Chabrier, G., Allard, F., \& Hauschildt, P.H. 1998, A\&A, 337, 403 (B98) 

\bibitem[Bessel (1990)]{bes90} Bessell, M. S. 1990, A\&AS, 83, 357

\bibitem[Bouvier et al.\ 1998]{bou98} Bouvier, J., Stauffer, J.R., 
Mart\'{\i}n, E.L., Barrado y Navascu\'es, D., Wallace, B., \& 
B\'ejar, V.J.S. 1998, A\&A, 336, 490

\bibitem[Burrows et al.\ 1993]{bur93} Burrows, A., Hubbard, W.B., 
Saumon, D., \& Lunine, J.I. 1993, ApJ, 406, 158 

\bibitem[Burrows et al.\ (1997)]{bur97} Burrows, A., Marley, M., 
Hubbard, W.B., Lunine, J.I., Guillot, T., Saumon, D., 
Freedman, R., Sudarsky, D., \& Sharp, C.  1997, 
ApJ, 491, 856

\bibitem[Calzetti \& Noll (1998)]{cal98} Calzetti, D., \& Noll, K. 1998, 
NICMOS Instrument Science Report, NICMOS-98-014

\bibitem[Canuto \& Mazzitelli (1991)]{can91} 
Canuto, V.M., \& Mazzitelli, I. 1991, ApJ, 370, 295

\bibitem[Cohen (1994)]{coh94}
Cohen M. 1994, AJ, 107, 582 

\bibitem[D'Antona \& Mazzitelli (1997)]{dm97}
D'Antona, F. \& Mazzitelli, I. 1997, Mem.S.A.It., 68, 807 (DM98)

\bibitem[de Zeeuw et al.\ 1999]{dez99}
de Zeeuw, P. T., Hoogerwerf, R., de Bruijne, J. H. J., 
Brown, A. G. A., \& Blaauw, A. 1999. 117, 354

\bibitem[Duchene et al.\ (1999)]{duc99} Duchene, G.,
Bouvier, J., \& Simon, T. 1999, A\&A, 343, 831 

\bibitem[Forrest et al.\ (1988)]{for88} Forrest, W. J., Shure, M., \&
Skrutskie, M. F. 1988, ApJ, 330, 119

\bibitem[Gliese (1969)]{gli69} Gliese, W. 1969, Catalogue of Nearby Stars,
Veroff. Heidelberg, No. 22

\bibitem[Greene \& Meyer 1995]{gre95} Greene, T. P. \& Meyer, M. R. 1995,
ApJ, 450, 233

\bibitem[Grossman et al.\ 1974]{gro74} Grossman, A.S., Hays, D., \& 
Graboske, H.C. 1974, A\&A, 30, 95 

\bibitem[Halbwachs et al.\ 2000]{hal00} Halbwachs, J.L., Arenou, F., 
Mayor, M., Udry, S., \& Queloz, D. 2000, A\&A, in press

\bibitem[Hartigan et al.\ 1989]{har89} Hartigan, P., Hartmann, L.,
Kenyon, S.J., Strom, S.E., Skrutskie, M.F.  1989, ApJS, 70, 899

\bibitem[Hartmann et al.\ (1991)]{har91} Hartmann, L., Stauffer, J. R.,
Kenyon, S. J., \& Jones, B. F. 1991, AJ, 101, 1050

\bibitem[Hauschildt et al.\ 1999]{hau99} Hauschildt, P. H., Peter, H., 
Allard, F., Fergusun, J., Baron, E., \& Alexander, D. R. 1999, ApJ, 525, 871

\bibitem[Henry \& Kirkpatrick (1990)]{hen90} Henry, T. J., \& Kirkpatrick,
J. D. 1990, ApJ, 354, 29

\bibitem[Henry et al.\ (1994)]{hen94} Henry, T. J., Kirkpatrick, J. D.,
\& Simons, D. A. 1994, AJ, 108, 1437

\bibitem[Herbig 1998]{her98} Herbig, G. H. 1998, ApJ, 497, 736

\bibitem[Hillenbrand 1997]{hil97} Hillenbrand, L.A. 1997, AJ, 113, 1733

\bibitem[Jones et al.\ 1994]{jon94} Jones, H. R. A., Longmore, A. J.,
Jameson, R. F., \& Mountain, C. M. 1994, MNRAS, 267, 413

\bibitem[Jones et al.\ 1995]{jon95} Jones, H. R. A., Longmore, A. J.,
Allard, F., Hauschildt, P. H., Miller, S., \& Tennyson, J. 1995, 
MNRAS, 277, 767

\bibitem[Jones et al.\ 1996]{jon96} Jones, H.R.A., Longmore, A.J., 
Allard, F., \& Hauschildt, P.H. 1996, MNRAS, 280, 77 

\bibitem[Keenan \& McNeil (1989)]{kee89} Keenan, P. C. \& McNeil, R. C.
1989, ApJS, 71, 245

\bibitem[Kenyon \& Hartmann (1995)]{ken95} Kenyon, S. J. \& Hartmann, L.
1995, ApJS, 101, 117

\bibitem[Kirkpatrick \& McCarthy 1994]{kir94} Kirkpatrick, J. D.,
\& McCarthy, D. W. Jr. 1994, AJ, 107, 333

\bibitem[Kirkpatrick et al.\ (1991)]{kir91} Kirkpatrick, J. D.,
Henry, T. J., \& McCarthy, D. W. Jr. 1991, ApJS, 77, 417

\bibitem[Kirkpatrick et al.\ (1995)]{kir95} Kirkpatrick, J. D.,
Henry, T. J., \& Simons, D. A. 1995, AJ, 109, 797

\bibitem[Krist \& Hook 1997]{kri97}Krist, J. \& Hook, R. 1997,
http://scivax.stsci.edu/$\sim$krist.tinytim.html

\bibitem[Lada \& Lada 1995]{lad95} Lada, E. A., \& Lada, C. J. 1995,
AJ, 109, 1682

\bibitem[Lasserre et al.\ (2000)]{las00} Lasserre, T., et al.\ 2000, 
A\&A, in press

\bibitem[Lauer (1999)]{lau99} Lauer, T. 1999, PASP, 111, 1434

\bibitem[Leggett 1992]{leg92} Leggett, S. K., 1992, ApJS, 82, 351

\bibitem[Leggett et al.\ 1996]{leg96} Leggett, S. K., Allard, F.,
Berriman, G., Dahn, C. C., \& Hauschildt, P. H. 1996, ApJS, 104, 117

\bibitem[LRLL]{luh98} Luhman, K. L., Rieke, G. H., 
Lada, C. J., \& Lada, E. A. 1998, ApJ, 508, 347 (LRLL)

\bibitem[Luhman (1999)]{luh99} Luhman, K.L. 1999, ApJ, 525, 466

\bibitem[Lynden-Bell \& Pringle (1974)]{lyn74}Lynden-Bell, D. \& 
Pringle, J. E. 1974, MNRAS, 168, 603

\bibitem[Marcy et al.\ 2000]{mar00}
Marcy, G.W., Cochran, W.D., \& Mayor, M. 2000, in 
``Protostars and Planets IV'', 
ed. V. Mannings, A. P. Boss, \& S. S. Russell 
(Tucson: University of Arizona Press), in press

\bibitem[Mathieu et al.\ 2000]{mat00}
Mathieu, R.D., Ghez, A.M., Jensen, E.L.N., \& Simon, M. 2000, 
in ``Protostars and Planets IV'', 
ed. V. Mannings, A. P. Boss, \& S. S. Russell 
(Tucson: University of Arizona Press), in press

\bibitem[Mayor et al.\ 1998]{may98}
Mayor, et al.\ 1998, in ``Brown Dwarfs and Extra-solar Planets'', 
ASP Conf. Ser. 134, 
eds. R. Rebolo, E. Mart\'{\i}n, \& M. R. Zapatero Osorio, p. 140

\bibitem[Meyer 1996]{mey96} Meyer, M. R. 1996, Ph.D. thesis, 
University of Massachusetts, Amherst

\bibitem[Meyer et al.\ 1997]{mey97} Meyer, M.R., Calvet, N., 
Hillenbrand, L.A. 1997, AJ, 114, 288

\bibitem[Meyer et al.\ (1998)]{mey98} Meyer, M.R., Edwards, S., Hinkle, K.H., 
\& Strom, S.E. 1998, ApJ, 508, 397

\bibitem[Meyer et al.\ 2000]{mey00} 
Meyer, M.R., Beckwith, S.V.W., Stauffer, J.R., \& Schultz, B. 2000,
        in preparation.

\bibitem[Miller \& Scalo (1979)]{mil79} Miller, G. E. \& Scalo, J. M. 1979,
ApJS, 41, 513

\bibitem[Perrin et al.\ 1998]{per98} Perrin, G., Coud\'e du Foresto, V., 
Ridgway, S.T., Mariotti, J.-M., Traub, W., Carleton, N.P., \& 
Lacasse, M.G. 1998, A\&A, 331, 619

\bibitem[Pirzkal \& Freudling 1998a]{pir98a} Pirzkal, N. \& Freudling, W.
1998a, ESA NICMOSlook User's Manual, revision 2.1

\bibitem[Pirzkal \& Freudling 1998b]{pir98b} Pirzkal, N. \& Freudling, W.
1998b, ESA CalnicC User's Manual, revision 2.1

\bibitem[Reid et al.\ 1999]{rei99} Reid, I.N., Kirkpatrick, J.D., 
Liebert, J., Burrows, A., Gizis, J.E., Burgasser, A., Dahn, C.C., 
Monet, D., Cutri, R., Beighman, C.A., \& Skrutskie, M. 1999, 
ApJ, 521, 613

\bibitem[Richichi et al.\ 1998]{ric98} Richichi, A., Fabbroni, L, 
Ragland, S., \& Scholz, M. 1999, A\&A, 344, 511

\bibitem[Rieke \& Lebofsky (1985)]{rie85} Rieke, G. H., \& 
Lebofsky, M. J. 1985, ApJ, 288, 618

\bibitem[Salpeter 1955]{sal55}Salpeter, E. E. 1955, ApJ, 121, 161

\bibitem[Saumon et al.\ 1996]{sau96} Saumon, D., Hubbard, W.B., 
Burrows, A., Guillot, T., Lunine, J.I., \& Chabrier, G. 1996, 
ApJ, 460, 993

\bibitem[Scalo (1986)]{sca86} Scalo, J. M. 1986, Fundam. Cosmic Phys., 11, 1

\bibitem[Scholz et al. (1999)]{sco99} Scholz, R.-D., 
Brunzendorf, J., Ivanov, G., Kharchenko, N., 
Lasker, B., Meusinger, H., Preibisch, T., 
Schilbach, E., \& Zinnecker, H. 1999, 
A\&A Supp., 137, 305

\bibitem[Shu 1995]{shu95} Shu, F.H. 1995, in 
``Molecular Clouds and Star Formation'', ed. C. Yuan
\& J. H. You (Singapore: World Scientific), 97-148

\bibitem[Silk 1977]{sil77} Silk, J. 1977, ApJ, 214, 152

\bibitem[Stauffer et al.\ 1995]{sta95} Stauffer, J. S., Hartmann, L. W., 
\& Barrado y Navascues, D. 1995, ApJ, 454, 910

\bibitem[Strom et al.\ (1989)]{str89} Strom, K. M., Strom, S. E.,
Edwards, S., Cabrit, S., \& Skrutskie, M. F. 1989, AJ, 97, 1451

\bibitem[Thompson et al.\ (1998)]{tho98} Thompson, R.I., Rieke, M., 
Schnieder, G., Hines, D.C., \& Corbin, M.R. 1998, ApJ, 492, L95

\bibitem[Tiede et al.\ 2000]{tie00}Tiede, G. P., Najita, J. R., \&
Carr, J. S. 2000, AJ, in preparation

\bibitem[Tinney et al.\ (1993)]{tin93} Tinney, C.G., Mould, J.R., \& 
Reid, I.N. 1993, AJ, 105, 1045

\bibitem[Tokunaga 1999]{tok99}Tokunaga, A. T. 1999, in Astrophysical
Quantities, edited by A. N. Cox (Berlin: Springer-Verlag)

\bibitem[Tsuji et al.\ (1996)]{tsu96}Tsuji, T., Ohnaka, K., \& Aoki, W. 1996, 
A\&A, 305, L1 

\bibitem[van Belle et al.\ 1999 ]{van99} van Belle, G.T. et al.\ 1999, 
AJ, 117, 521

\bibitem[Wainscoat et al.\ 1992]{wai92} Wainscoat, R.J., Martin, C., 
Volk, K., Walker, H.J., \& Schwartz, D.E. 1992, ApJS, 83, 111

\bibitem[Walter et al.\ (1988)]{wal88} Walter, F. M., Brown, A.,
Mathiew, R. D., Myers, P. C., \& Vrba, F. J. 1988, AJ, 96, 297

\bibitem[White et al.\ 1999]{whi99} White, R.J., Ghez, A.M., 
Reid, I.N., \& Schultz, G. 1999, ApJ, 520, 811

\bibitem[Wilking et al.\ 1999]{wil99} Wilking, B.A., Greene, T.P., 
\& Meyer, M.R. 1999, AJ, 117, 469

\bibitem[Zapatero Osorio et al.\ 2000]{zap00} Zapatero Osorio, M. R., 
B\'ejar, V. J. S., Rebolo, R., Mart\'{\i}n, E. L., \& Basri, G. 2000, 
ApJ Letters, in press

\end{thebibliography}
\end{document}